\documentclass[preprintnumbers,amsmath,amssymb,showpacs,nofootinbib]{revtex4}
\usepackage{marvosym,pslatex}
\usepackage{graphicx,amsmath,amsfonts,amssymb,dcolumn,amsthm}
\usepackage{slashed}
\usepackage{bbm}                
\usepackage{xspace}
\usepackage{pgffor}	
\usepackage{tikz}
\usepackage{amssymb}
\usepackage{amsmath}
\usepackage{amsfonts}
\usepackage{mathbbol}
\usepackage{mathrsfs}
\usepackage{amssymb}
\usepackage{latexsym}

\begin{document}

\title{Renormalizability of Yang-Mills theory with Lorentz violation and gluon mass generation}

\author{T.~R.~S.~Santos}
\email{tiagoribeiro@if.uff.br}
\affiliation{UFF $-$ Universidade Federal Fluminense, Instituto de F\'{\i}sica, Campus da Praia Vermelha, Avenida General Milton Tavares de Souza s/n, 24210-346, Niter\'oi, RJ, Brasil.}
\author{R.~F.~Sobreiro}
\email{sobreiro@if.uff.br}
\affiliation{UFF $-$ Universidade Federal Fluminense, Instituto de F\'{\i}sica, Campus da Praia Vermelha, Avenida General Milton Tavares de Souza s/n, 24210-346, Niter\'oi, RJ, Brasil.}

\begin{abstract}
We show that pure Yang-Mills theories with Lorentz violation are renormalizable to all orders in perturbation theory. To do this, we employ the algebraic renormalization technique. Specifically, we control the breaking terms with a suitable set of external sources which, eventually, attain certain physical values. The Abelian case is also analyzed as a starting point. The main result is that the renormalizability of the usual Maxwell and Yang-Mills sectores are both left unchanged. Furthermore, in contrast to Lorentz violating QED, the odd CPT violation sector of Yang-Mills theories renormalizes independently. Moreover, the method induces, in a natural way, mass terms for the gauge field while the photon remains massless (at least n the sense of a Proca-like term). The entire analysis is carried out at the Landau gauge.
\end{abstract}

\pacs{11.10.Gh, 11.15.-q, 11.30.Cp, 12.60.-i}

\maketitle

\section{Introduction}\label{intro}

Lorentz and gauge symmetries play an important and, perhaps, indispensable role in quantum field theory and particle Physics \cite{Bargmann:1946me,Bargmann:1948ck,Becchi:1975nq,Tyutin:1975qk}. From the classification of particles to renormalizability proofs, these symmetries are crucial. However, theories for which Lorentz symmetry is not required have been receiving considerable attention in the last decades \cite{Colladay:1996iz,Colladay:1998fq,Jackiw:1999yp,Kostelecky:2003fs,Kostelecky:2005ic,Diaz:2011ia}. Although direct effects of such theories would appear only beyond Planck scale, some ``cumulative'' effects could arise \cite{Bertolami:1996cq,Bluhm:1997ci,Bailey:2006fd,Kostelecky:2013rv}. Even though this type of theories originates as effective models from an extremely high energy theory \cite{Kostelecky:1988zi,Altschul:2005mu}, it should be studied in the context of quantum field theory. And, in order to provide reliable and consistent theoretical predictions, attributes as stability, renormalizability, unitarity and causality are very welcome features. For example, stability requires that the Hamiltonian of the theory is bounded from below, and causality refers to commutativity of observables at space-like intervals, see for instance \cite{Colladay:1996iz,Colladay:1998fq,Adam:2001ma,Adam:2001kx,Adam:2002rg} for more details. In this work we confine ourselves to a detailed analysis of the renormalizability of pure non-Abelian gauge theories with Lorentz violation.

Models with Lorentz and CPT broken symmetries are characterized by the presence of background tensorial fields coupled to the fundamental fields of the theory. Typically, the Lorentz violation background fields arise at the scenario of effective field theories originated from fundamental models such as string theories \cite{Kostelecky:1988zi}, non-commutative field theories \cite{Guralnik:2001ax, Carroll:2001ws, Carmona:2002iv, Carlson:2002zb, SheikhJabbari:2000vi}, supersymmetric field theories \cite{Berger:2003ay, Berger:2001rm, GrootNibbelink:2004za} and loop quantum gravity \cite{Alfaro:2004ur}. In string theory, for instance, the Lorentz symmetry breaking arises from a spontaneous symmetry breaking. Specifically, from non-trivial vacuum value expectation of the tensorial fields. Such background fields could contain effects of an underlying fundamental theory at the Planck mass scale $M_{P}\sim10^{19} GeV$. In fact, there exist some expectation to detect possible signals for bounds of these violating coefficients, such as in high precision experiments in atomics processes \cite{Bluhm:1997ci,Bear:2000cd,Bluhm:1999ev,Bluhm:1998rk}. A theoretical proposal to describe the Lorentz symmetry breaking in this scale is the standard model extension (SME). At this model, the Lorentz breaking coefficients are introduced through couplings with fundamentals fields of the standard model and the model is power-counting renormalizable \cite{Colladay:1998fq}. Another theoretical proposal for Lorentz violation is the modified dispersion relations (MDR) \cite{Lehnert:2003ue}. Essentially, these new dispersion relations carry extra contributions that depends on the energy scale and are only meaningful at ultra high energies, being suppressed at the low energy limit. In principle, ultra high energy cosmic rays at the Planck energy scale where Lorentz and CPT symmetry breaking would take place is encoded in astrophysical processes. A possible explanation to the observation of the apparent excess of cosmic rays in this region of energy \cite{Coleman:1998ti} is the MDR which, in this case, suggest that these cosmic rays could develop velocities faster than light velocity. Concerning the renormalization properties of Lorentz violating QED, a 1-loop renormalization analysis was already discussed in \cite{Kostelecky:2001jc} and a full algebraic study at all orders in perturbation theory was established in \cite{DelCima:2012gb}. Another interesting study about renormalizability issues on Lorentz and CPT violating QED was performed in \cite{deBerredoPeixoto:2006wz}. In that work, it was assumed which the fields of this model reside in a curved manifold, and the Lorentz and CPT violating parameters are treated as classical fields rather than constants, which happens to be very similar to the approach employed in the present work.

The non-Abelian sector of the standard model extension, till now, has received few attention from both, theoretical studies and experimental tests for the bounds of the Lorentz violating background parameters. As pointed out in \cite{Colladay:2006rk}, the ultraviolet behavior of the even CPT coupling may give a great bound for this coefficients, in contrast to odd CPT couplings. In what concerns the renormalization properties of pure Yang-Mills theory with Lorentz violation, it was shown in \cite{Colladay:2006rk}, that this model can be renormalized at one-loop order. It is worth mention that a non-Abelian Chern-Simons-like term can be induced from the Abelian Lorentz violating term at 1-loop radiative corrections \cite{Gomes:2007rv}.

In the present work we focus our study on the non-Abelian sector of the standard model extension (SME), \textit{i.e.}, pure Yang-Mills theory with Lorentz violation. In particular, we employ the algebraic renormalization approach \cite{Piguet:1995er} to prove that this model is renormalizable, at least to all orders in perturbation theory. In our analysis we include all possible breaking terms. Besides BRST quantization, we introduce a suitable set of sources that controls the Lorentz breaking terms. Eventually, in order to regain the original action, these sources attain specific physical values. This trick is originally due to Symanzik \cite{Symanzik:1969ek} and was vastly employed in non-Abelian gauge theories in order to control a soft BRST symmetry breaking, see for instance \cite{Zwanziger:1992qr,Dudal:2005na,Baulieu:2008fy,Baulieu:2009xr,Dudal:2011gd}. Essentially, the broken model is embedded into a larger theory where the relevant symmetry is respected. Then, after renormalization, the theory is contracted down to the original model. We will give attention, firstly, to the Abelian theory in the presence of Lorentz violation and in the absence of fermions\footnote{In fact the presence of fermions in a Lorentz violating model, even in a Abelian model, will make the study of the renormalizability very difficult, at least by our approach. Thus the Abelian model is studied here in the absence of fermions in order to compare it with the non-Abelian case; the later introduces many difficulties compared with the former, even in the absence of fermions. The study of the fermionic sector is left for future investigation \cite{S}.}. Adopting the Symanzik source approach, we can introduce the most general action which carries, for instance, vacuum type terms as well as dimension two condensate terms. The price we pay is that extra independent renormalizations parameters are needed to account for the extra vacuum divergences. Remarkably, the extra condensate type term $A^a_{\mu}A^a_{\mu}$ arises due a coupling with the odd CPT sector of the model, also with an independent renormalization coefficient. We have then an induced mass term for the gluon originating from the Lorentz violating terms. However, these terms are rule out in the Lorentz violating Maxwell's theory due the fact that the ghost equation is not integrated, making it stronger than its non-Abelian version. These different characteristics between the Ward identities of the Abelian and non-Abelian models will result in different renormalization properties among Maxwell and Yang-Mills Lorentz violation coefficients. For instance, we will show that the odd CPT breaking term at the Maxwell theory, $\epsilon_{\mu\nu\alpha\beta}v_{\mu}A_{\nu}\partial_{\alpha}A_{\beta}$, does not renormalize. Nonetheless, the odd CPT breaking term in Yang-Mills theory renormalizes independently.

This work is organized as follows: Sect.~\ref{QED} is dedicated to the renormalizability proof of the Maxwell theory with Lorentz violation. In Sect.~\ref{YM}, we provide the definitions and conventions of the pure Yang-Mills theory with Lorentz violation and the BRST quantization of the model with the extra set of auxiliary sources is provided. Then, in Sect.~\ref{RENORMALIZABILITY}, we study the renormalizability of the model. Our final considerations are displayed in Sect.~\ref{FINAL}.

\section{Lorentz violating Maxwell theory}\label{QED}

We consider the $U(1)$ Abelian gauge theory with Lorentz violation. For mere convenience, the scenario for this theory (and also for the non-Abelian case) is the Euclidean four-dimensional spacetime\footnote{Besides the fact that Euclidean metric is simpler handle, this choice is convenient in the treatment of non-perturbative effects where it is unknown if Wick's rotation is valid.}. The action of the model is the following\footnote{We are not considering fermions in this work, as mentioned at the Introduction.} \cite{Kostelecky:2001jc}
\begin{eqnarray}
S_0&=&S_{M}+S_{LVE}+S_{LVO}\;,
\label{A}
\end{eqnarray}
where
\begin{eqnarray}
S_{M}&=&\frac{1}{4}\int d^4xF_{\mu\nu}F_{\mu\nu}\;,
\label{A1}
\end{eqnarray}
is Maxwell's action. The field strength is defined as $F_{\mu\nu}=\partial_{\mu}A_{\nu}-\partial_{\nu}A_{\mu}$, with $A_\mu$ being the gauge field. The Lorentz violating sector of even CPT is given by
\begin{eqnarray}
S_{LVE}&=&\frac{1}{4}\int d^4x\kappa_{\alpha\beta\mu\nu}F_{\alpha\beta}F_{\mu\nu}\;,
\label{A3}
\end{eqnarray}
while the odd CPT Lorentz violation term is defined as
\begin{eqnarray}
S_{LVO}&=&\int d^4x\epsilon_{\mu\nu\alpha\beta}v_{\mu}A_{\nu}\partial_{\alpha}A_{\beta}\;.
\label{A4}
\end{eqnarray}
The Lorentz violation is characterized by the fields $v_{\mu}$, with mass dimension 1, and $\kappa_{\alpha\beta\mu\nu}$, which is dimensionless. These tensors fix privileged directions in spacetime, dooming it to anisotropy. Tensorial fields with even numbers of indices preserve CPT while tensors with odd number of indices do not. The tensor $\kappa_{\alpha\beta\mu\nu}$ obeys the same properties of the Riemann tensor, and is double traceless:
\begin{align}
&\kappa_{\alpha\beta\mu\nu}\;=\;\kappa_{\mu\nu\alpha\beta}\;=\;-\kappa_{\beta\alpha\mu\nu}\ ,\nonumber\\
&\kappa_{\alpha\beta\mu\nu}+\kappa_{\alpha\mu\nu\beta}+\kappa_{\alpha\nu\beta\mu}\;=\;0\ ,\nonumber\\
&\kappa^{\mu\nu}_{\;\;\; \mu\nu}\;=\;0\ .
\label{3}
\end{align}
As the reader can easily verify, the action \eqref{A} is a Lorentz scalar, being invariant under observers Lorentz transformations while, in contrast, presents violation with respect to particle Lorentz transformations.

In the present work we employ the BRST quantization method and adopt the Landau gauge condition $\partial_\mu A_{\mu}=0$. Thus, besides the photon field, we introduce the Lautrup-Nakanishi field $b$ and the Faddeev-Popov ghost and anti-ghost fields, namely, $c$ and $\overline{c}$, respectively. The respective BRST transformations are
\begin{eqnarray}
sA_{\mu}&=&-\partial_{\mu}c\;,\nonumber\\
sc&=&0\;,\nonumber\\
s\bar{c}&=&b\;,\nonumber\\
sb&=&0\;,
\label{A5}
\end{eqnarray}
where $s$ is the nilpotent BRST operator. The quantum numbers of the fields and background tensors are displayed in table \ref{tableA1}. The full Landau gauge fixed action is
\begin{eqnarray}
S_0&=&S_{M}+S_{LVE}+S_{LVO}+S_{gf}\;,
\label{A6}
\end{eqnarray}
where 
\begin{eqnarray}
S_{gf}&=&s\int d^4x\overline{c}\partial_{\mu}A_{\mu}\;=\;\int d^4x\left(b\partial_{\mu}A_{\mu}+\overline{c}\partial^2c\right)\;,
\label{A7}
\end{eqnarray}
is the gauge fixing action enforcing the Landau gauge condition. The Landau gauge is chosen due to a few simple reasons \cite{Piguet:1995er}: i.) It is a covariant gauge; ii.) It has a rich content of symmetries; iii.) It is a fixed point of the renormalization group; iv) It is the simplest case, so it is a convenient starting choice; v.) It is renormalizable in the ordinary case. 
\begin{table}[h]
\centering
\begin{tabular}{|c|c|c|c|c|c|c|}
	\hline 
fields/tensors & $A$ & $b$ &$c$ & $\bar{c}$ & $v$ & $\kappa$ \\
	\hline 
UV dimension & $1$ & $2$ & $0$ & $2$ & $1$ & $0$ \\ 
Ghost number & $0$ & $0$ & $1$& $-1$ & $0$ & $0$\\ 
\hline 
\end{tabular}
\caption{Quantum numbers of the fields and background tensors.}
\label{tableA1}
\end{table}

Lorentz symmetry plays a fundamental role on the renormalizability of gauge theories, thus, the presence of a Lorentz violating sector demands extra care. To deal with this obstacle we replace each of the background tensors by an external classical source and, possibly, its BRST doublet counterpart (if needed). Thus, the local composite operator whose each background tensor is a coefficient, will appear as coupled to one of these sources. Indeed, there will be two classes of sources: BRST invariant sources and BRST doublet sources. The first class will be coupled to the BRST/gauge invariant composite operators while the second class couples to the other operators. Thus, we define the following  invariant source
\begin{eqnarray}
s\bar{\kappa}_{\alpha\beta\mu\nu}&=&0\;.
\label{A8}
\end{eqnarray}
And the BRST doublet sources are given by
\begin{eqnarray}
s\lambda_{\mu\nu\alpha}&=&J_{\mu\nu\alpha}\;, \nonumber\\
sJ_{\mu\nu\alpha}&=&0\;.
\label{A9}
\end{eqnarray}
The quantum numbers of the sources are displayed in table \ref{tableA3}. Eventually, these sources will attain the following physical values
\begin{eqnarray}
J_{\mu\nu\alpha}\mid_{phys}&=&v_{\beta}\epsilon_{\beta\mu\nu\alpha}\;,\nonumber\\
\lambda_{\mu\nu\alpha}\mid_{phys}&=&0\;,\nonumber\\
\bar{\kappa}_{\alpha\beta\mu\nu}\mid_{phys}&=&\kappa_{\alpha\beta\mu\nu}\;.
\label{A10}
\end{eqnarray}
\begin{table}[h]
\centering
\begin{tabular}{|c|c|c|c|}
	\hline 
sources & $\lambda$ & $J$ & $\bar{\kappa}$  \\
	\hline 
UV dimension & $1$ & $1$ & $0$ \\ 
Ghost number &  $-1$ & $0$ & $0$ \\ 
\hline 
\end{tabular}
\caption{Quantum numbers of the sources.}
\label{tableA3}
\end{table}
Thus, we replace the action \eqref{A6} by\footnote{Since the Lorentz breaking is now controlled by the external sources, we rename the original actions without the letter "V", of violation.}
\begin{eqnarray}
S&=&S_{M}+S_{LO}+S_{LE}+S_{gf}\ ,
\label{B1}
\end{eqnarray}
where, now,
\begin{eqnarray}
S_{LE}&=&\frac{1}{4}\int d^4x\bar{\kappa}_{\alpha\beta\mu\nu}F_{\alpha\beta}F_{\mu\nu}\;,\nonumber\\
S_{LO}&=&s\int d^4x\lambda_{\mu\nu\alpha}A_{\mu}\partial_{\nu}A_{\alpha}\;=\;\int d^4x\left(J_{\mu\nu\alpha}A_{\mu}\partial_{\nu}A_{\alpha}+\lambda_{\mu\nu\alpha}\partial_{\mu}c\partial_{\nu}A_{\alpha}\right)\;.
\label{B2}
\end{eqnarray}
is the embedding of the Lorentz violating bosonic sector. The BRST symmetry demands that all possible terms, \textit{i.e.}, integrated local polynomials in the fields and sources with dimension four and vanishing ghost number, that respect BRST symmetry, must be added to the model. Then, through the algebraic renormalization techniques, the Ward identities will select the terms that are actually needed (see next Section). Power counting renormalizability also allows one more term to be added to the action \eqref{B1}, namely
\begin{eqnarray}
S_V&=&s\int d^4x\left(\zeta \lambda_{\mu\nu\alpha}J_{\mu\beta\gamma}J_{\nu\beta\kappa}J_{\gamma\kappa\alpha}+\vartheta\bar{\kappa}_{\mu\nu\alpha\beta}\lambda_{\mu\rho\omega}J_{\nu\rho\sigma}J_{\alpha\omega\delta}J_{\beta\sigma\delta}\right)\nonumber\\
&=&\int d^4x\left(\zeta J_{\mu\nu\alpha}J_{\mu\beta\gamma}J_{\nu\beta\kappa}J_{\gamma\kappa\alpha}+\vartheta\bar{\kappa}_{\mu\nu\alpha\beta}J_{\mu\rho\omega}J_{\nu\rho\sigma}J_{\alpha\omega\delta}J_{\beta\sigma\delta}\right)\;.
\label{B3}
\end{eqnarray}
The dimensionless parameters $\zeta$ and $\vartheta$ are introduced to absorb possible vacuum divergences. A remark must be made at this point: In principle, from power counting analysis, we could add a series of terms of the type $\bar{\kappa}_{\alpha\beta\rho\sigma}\bar{\kappa}_{\rho\sigma\mu\nu}F^a_{\alpha\beta}F^a_{\mu\nu}$, $\bar{\kappa}_{\alpha\beta\rho\sigma}\bar{\kappa}_{\rho\sigma\omega\delta}\bar{\kappa}_{\omega\delta\mu\nu}F^a_{\alpha\beta}F^a_{\mu\nu}$ and so on. Nonetheless, all these terms could be rearranged in only one term coupled to the operator $F^a_{\alpha\beta}F^a_{\mu\nu}$. This infinite series can then be renamed as a single source term by means of the first of \eqref{B2}, preserving the original term. In fact, this argument is valid for all terms that mix with $\bar{\kappa}_{\mu\nu\alpha\beta}$ in the Abelian or non-Abelian cases. Formally, one can consider the infinite tower of terms, and the respective counterterms, and only after the absorption of the divergences the redefinition applies. Obviously, the classical character of $\bar{\kappa}_{\mu\nu\alpha\beta}$ is crucial to this argument. See also \cite{Netto:2014faa,Shapiro:2013bsa}.

The complete action we have is
\begin{eqnarray}
\Xi&=&S+S_V\;.
\label{B4}
\end{eqnarray}
Explicitly, the action \eqref{B4} has the following form
\begin{eqnarray}
\Xi&=&\frac{1}{4}\int d^4xF_{\mu\nu}F_{\mu\nu}+\frac{1}{4}\int d^4x\bar{\kappa}_{\alpha\beta\mu\nu}F_{\alpha\beta}F_{\mu\nu}+\int d^4x\left(J_{\mu\nu\alpha}A_{\mu}\partial_{\nu}A_\alpha+\lambda_{\mu\nu\alpha}\partial_{\mu}c\partial_{\nu}A_{\alpha}\right)+\nonumber\\
&+&\int d^4x\left(b\partial_{\mu}A_{\mu}+\bar{c}\partial^2 c\right)+\int d^4x\left(\zeta J_{\mu\nu\alpha}J_{\mu\beta\gamma}J_{\nu\beta\kappa}J_{\gamma\kappa\alpha}+\vartheta\bar{\kappa}_{\mu\nu\alpha\beta}J_{\mu\rho\omega}J_{\nu\rho\sigma}J_{\alpha\omega\delta}J_{\beta\sigma\delta}\right)\;.\nonumber\\
\label{B5}
\end{eqnarray}
The action \eqref{B5}, at the physical value of the sources \eqref{A10}, reduces to
\begin{eqnarray}
\Xi_{phys}&=&\frac{1}{4}\int d^4xF_{\mu\nu}F_{\mu\nu}+\frac{1}{4}\int d^4x\kappa_{\alpha\beta\mu\nu}F_{\alpha\beta}F_{\mu\nu}+\int d^4xv_{\beta}\epsilon_{\beta\mu\nu\alpha}A_{\mu}\partial_{\nu}A_{\alpha}+\nonumber\\
&+&\int d^4x\left(b\partial_{\mu}A_{\mu}+\bar{c}\partial^2c\right)+2v^2\int d^4x\left(3\zeta v^2-\vartheta\kappa_{\alpha\mu\sigma\mu}v_{\alpha}v_{\sigma}\right)\:.
\label{B6}
\end{eqnarray}

A remark is in order now: The source $J$ is introduced as a BRST doublet where its BRST counterpart is the source $\lambda$. As a consequence, the entire term depending on $J$ and $\lambda$ is an exact BRST variation. Thus, it belongs to the nonphysical sector of the model. However, the model suffers a contraction in order to be deformed to the action of interest (the physical action). Under such contraction, this term is thrown to the physical sector of the theory. In fact, the terms depending on $v_\mu$, at the physical action, cannot be written as a BRST exact variation anymore. Let us put this in other words. The physical action \eqref{B6} is the true action, violates Lorentz symmetry and the violating terms cannot be written as a BRST exact variation. Thus, in order to study its renormalizability, the theory is embedded into a larger theory which displays full Lorentz and BRST symmetries. The embedding is characterized by the auxiliary sources which enter in the place of the violating parameters. The physical theory is recovered from a specific choice of these sources (the physical values). These values are attained by contracting the functional space of the sources into the $\mathbb{R}^4$-space of the vector $v_\mu$. The main idea of the method is that the model is renormalized in its embedded form and only after the renormalization the model is contracted to the physical sector.

For completeness, we compute the propagator for the photon, at the Landau gauge, taking\footnote{The presence of a general $\kappa_{\alpha\beta\mu\nu}$ makes the computation highly non-trivial. For a detailed study on this sector see, for instance, \cite{Casana:2009xs}.} $\kappa_{\alpha\beta\mu\nu}=0$. The result is
\begin{eqnarray}
\langle A_{\mu}(k)A_{\nu}(-k)\rangle&=&\frac{1}{Q}\left[k^2\theta_{\mu\nu}-\frac{4(v_\alpha k_\alpha)^2}{k^2}\omega_{\mu\nu}+2S_{\mu\nu}+\frac{4(v_\alpha k_\alpha)}{k^2}\Sigma_{\mu\nu}-4\Lambda_{\mu\nu}\right]\;,
\label{B6A}
\end{eqnarray}
where $Q=k^4-4[v^2k^2-(v_\alpha k_\alpha)^2]$, and the operators
\begin{eqnarray}
\theta_{\mu\nu}&=&\delta_{\mu\nu}-\frac{k_{\mu}k_{\nu}}{k^2}\;,\nonumber\\	
\omega_{\mu\nu}&=&\frac{k_{\mu}k_{\nu}}{k^2}\;,\nonumber\\
S_{\mu\nu}&=&i\epsilon_{\mu\nu\alpha\beta}v_{\alpha}k_{\beta}\;,\nonumber\\
\Sigma_{\mu\nu}&=&v_{\mu}k_{\nu}+v_{\nu}k_{\mu}\;,\nonumber\\
\Lambda_{\mu\nu}&=&v_{\mu}v_{\nu}\;,
\label{B6B}
\end{eqnarray}
form a closed algebra. See, for instance, \cite{BaetaScarpelli:2003yd} for more details. It is worth to mention here that the physical modes of the gauge field, \textit{i.e.}, the photon, do not change with respect to the usual Maxwell theory with Lorentz violation; our approach does not change the kinetic part of this model and does not generate any Proca-like terms. Thus, the causality and unitarity of the model are maintained \cite{Adam:2001ma}. However, as it is clear from \eqref{B6}, the vacuum of the model changes when we take the physical limit of the sources in the action \eqref{B3}. We will discuss these points again at the non-Abelian case.

\subsection{Renormalizability}

In order to proof that this model is renormalizable to all orders in perturbation theory, let us now display the full set of Ward identities obeyed by the action \eqref{B5}:
\begin{itemize}
	\item Slavnov-Taylor identity
	\begin{eqnarray}
\mathcal{S}(\Xi)&=&\int d^4x\left(-\partial_{\mu} c\frac{\delta \Xi}{\delta A_{\mu}}+b\frac{\delta \Xi}{\delta \overline{c}}+J_{\mu\nu\alpha}\frac{\delta \Xi}{\delta\lambda_{\mu\nu\alpha}}\right)=0\;.
\label{B7}
\end{eqnarray}

\item Gauge fixing and anti-ghost equations
\begin{eqnarray}
\frac{\delta \Xi}{\delta b}&=&\partial_{\mu}A_{\mu}\;,\nonumber\\
\frac{\delta \Xi}{\delta \overline{c}}&=&\partial^2c\;.
\label{B8}
\end{eqnarray}

\item Ghost equation
\begin{eqnarray}
\frac{\delta \Xi}{\delta c}&=&\partial_\mu\left(\lambda_{\mu\nu\alpha}\partial_\nu A_\alpha\right)-\partial^2\overline{c}\;.
\label{B9}
\end{eqnarray}

\end{itemize}
At \eqref{B8} and \eqref{B9}, the breaking terms are linear in the fields. Thus it will remain at classical level \cite{Piguet:1995er}. From \eqref{B9} it is possible to predict that the odd-CPT Lorentz violating sector of the Maxwell theory will not suffer renormalization. This is due to the fact that this term induces a violation of ghost equation. As a consequence, a counterterm associated with the odd-CPT Lorentz violating will be eliminated by the Ward identity \eqref{B9}. 

In order to obtain the most general counterterm which can be freely added to the classical action $\Xi$ at any order in perturbation theory, we define a general local integrated polynomial $\Xi^c$ with dimension bounded by four and vanishing ghost number. Thus, imposing the Ward identities (\ref{B7}-\ref{B9}) to the perturbed action $\Xi+\varepsilon \Xi^c$, where $\varepsilon$ is a small parameter, it is easy to find that the counterterm must obey the following constraints
\begin{eqnarray}
\mathcal{B}_{\Xi}\Xi^c&=&0\;,\nonumber\\
\frac{\delta \Xi^c}{\delta b}&=&0\;,\nonumber\\
\frac{\delta \Xi^c}{\delta \overline{c}}&=&0\;,\nonumber\\
\frac{\delta \Xi^c}{\delta c}&=&0\;,
\label{B10}
\end{eqnarray}
where the operator $\mathcal{B}_{\Xi}$ is the nilpotent Slavnov-Taylor operator,
\begin{eqnarray}
\mathcal{B}_{\Xi}=\int d^4x\left(-\partial_{\mu} c\frac{\delta}{\delta A_{\mu}}+b\frac{\delta}{\delta\bar{c}}+J_{\mu\nu\alpha}\frac{\delta }{\delta\lambda_{\mu\nu\alpha}}\right)\;.
\label{C1}
\end{eqnarray}
The first constraint of \eqref{B10} states that to find the invariant counterterm is a cohomology problem for the operator $\mathcal{B}_{\Xi}$ in the space of the integrated local field polynomials of dimension four. From the general results on algebraic renormalization \cite{Piguet:1995er}, it is an easy task to find
\begin{eqnarray}
\Xi^c&=&\frac{1}{4}\int d^4x\;a_0F_{\mu\nu}F_{\mu\nu}+\frac{1}{4}\int d^4x\;a_1\bar{\kappa}_{\alpha\beta\mu\nu}F_{\alpha\beta}F_{\mu\nu}+\mathcal{B}_{\Xi}\Delta^{(-1)}\;,
\label{C2}
\end{eqnarray}
where $\Delta^{(-1)}$ is the most general local polynomial counterterm with dimension bounded by four and ghost number $-1$, given by
\begin{eqnarray}
\Delta^{(-1)}&=&\int d^4x\;\left(a_2\bar{c}\partial_\mu A_\mu +a_3\overline{c}b+a_4\lambda_{\mu\nu\alpha}A_{\mu}\partial_{\nu}A_{\alpha}+a_5\zeta\lambda_{\mu\nu\alpha}J_{\mu\beta\gamma}J_{\nu\beta\kappa}J_{\gamma\kappa\alpha}+\right.\nonumber\\
&+&\left.a_6\vartheta\bar{\kappa}_{\mu\nu\alpha\beta}\lambda_{\mu\rho\omega}J_{\nu\rho\sigma}J_{\alpha\omega\delta}J_{\beta\sigma\delta}\right)\;,
\label{C3}
\end{eqnarray}
where the parameters $a_i$ are free coefficients. Defining $\hat{\Xi}=\mathcal{B}_{\Xi}\Delta^{(-1)}$ one finds
\begin{eqnarray}
\hat{\Xi}&=&a_2\int d^4x\left(b\partial_{\mu}A_{\mu}+\overline{c}\partial^2c\right)+a_3\int d^4xb^2+a_4\int d^4x\left(J_{\mu\nu\alpha}A_{\mu}\partial_{\nu}A_{\alpha}+\lambda_{\mu\nu\alpha}\partial_{\mu}c\partial_{\nu}A_{\alpha}\right)+\nonumber\\
&+&a_5\int d^4x\zeta J_{\mu\nu\alpha}J_{\mu\beta\gamma}J_{\nu\beta\kappa}J_{\gamma\kappa\alpha}+a_6\int d^4 x\vartheta\bar{\kappa}_{\mu\nu\alpha\beta}J_{\mu\rho\omega}J_{\nu\rho\sigma}J_{\alpha\omega\delta}J_{\beta\sigma\delta}\;.
\label{C4}
\end{eqnarray}
From the second or third constraints in \eqref{B10}, it follows that $a_2=a_3=0$. Moreover, from the ghost equation, $a_4=0$. It follows then that the most general counterterm allowed by the Ward identities is given by
\begin{eqnarray}
\Xi^c&=&\frac{1}{4}\int d^4x\;a_0F_{\mu\nu}F_{\mu\nu}+\frac{1}{4}\int d^4x\;a_1\bar{\kappa}_{\alpha\beta\mu\nu}F_{\alpha\beta}F_{\mu\nu}+a_5\int d^4x\zeta J_{\mu\nu\alpha}J_{\mu\beta\gamma}J_{\nu\beta\kappa}J_{\gamma\kappa\alpha}+\nonumber\\
&+&a_6\int d^4 x\vartheta\bar{\kappa}_{\mu\nu\alpha\beta}J_{\mu\rho\omega}J_{\nu\rho\sigma}J_{\alpha\omega\delta}J_{\beta\sigma\delta}\;.
\label{C5}
\end{eqnarray}

It remains to infer if the counterterm $\Xi^c$ can be reabsorbed by the original action $\Xi$ by means of the redefinition of the fields, sources and parameters of the theory through
\begin{eqnarray}
\Xi(\Phi, J, \xi)+\varepsilon \Xi^c(\Phi, J, \xi)&=& \Xi(\Phi_0, J_0, \xi_0)+\mathcal{O}(\varepsilon^2)\;,
\label{C6}
\end{eqnarray}
where the bare fields, sources and parameters are defined as
\begin{eqnarray}
\Phi_0&=&Z^{1/2}_{\Phi}\Phi\;,\;\;\;\;\Phi \in \left\{A, b, \bar{c}, c\right\}\;,\nonumber\\
J_0&=&Z_JJ\;,\;\;\;\;\;\;\;J \in \left\{J,\lambda,\bar{\kappa}\right\}\;, \nonumber\\
\xi_0&=&Z_{\xi}\xi\;,\;\;\;\;\;\;\;\;\;\xi \in \left\{\vartheta,\zeta\right\}\;.
\label{C7}
\end{eqnarray}
It is not difficult to check that this can be performed, providing the multiplicative renormalizability proof of the theory to all orders in perturbation theory. In fact, for the independent renormalization factor of the  photon, we have
\begin{eqnarray}
Z_A^{1/2}&=&1+\frac{1}{2}\varepsilon a_0\;.
\label{renC1}
\end{eqnarray}
The  ghost fields does not renormalize,
\begin{eqnarray}
Z_c^{1/2}&=&Z_{\overline{c}}^{1/2}\;\;=\;\;1\;,
\label{renC2}
\end{eqnarray}
and the Lautrup-Nakanishi field renormalization is not independent
\begin{eqnarray}
Z_b^{1/2}&=&Z_A^{-1/2}\;.
\label{renC3}
\end{eqnarray}
Thus, the standard QED sector remains unchanged with respect to the ordinary case. For the violating sector we have
\begin{eqnarray}
Z_{J}&=&Z^2_\lambda\;\;=\;\;Z_A^{-1}\;,\nonumber\\
Z_\zeta&=&1+\varepsilon\left(a_5+4a_0\right)\;,\nonumber\\
Z_{\bar{\kappa}}&=&1+\varepsilon(a_1-a_0)\;,\nonumber\\
Z_{\vartheta}&=&1+\varepsilon(a_6-a_1+5a_0)\;.
\label{renC4}
\end{eqnarray}
From the first equation in \eqref{renC4}, as we have pointed out before, we see that the odd-CPT Lorentz violating coefficient $v_{\mu}$ does not renormalizes independently, namely, its renormalization depends only of the photon renormalization. This is also clear from the final counterterm \eqref{C5}, where the CPT-odd part is not present and, thus, does not renormalize. This ends the renormalizability of the Lorentz violating Abelian gauge theory, at least to all orders in perturbation theory.

The study of the renormalizability of pure QED might be seen as an unnecessary effort since the theory is free (we are not considering fermions at this point). In fact, no interaction terms would be generated from the analysis of quantum stability and no parameters would be renormalized, only the fields would. Nevertheless, the study of the quantum stability of Maxwell theory with Lorentz violation under the method of external auxiliary sources can establish if the model accepts or not other quadratic terms involving the sources (for instance, a mass term of the type $v^2 A_\mu A_\mu$ could appear at the physical limit). Thus, the study of the free Abelian case can be used as a first consistency check of the method. Nevertheless, the presence of the quartic $J$-source terms generate independent renormalizations of the vacuum energy. Moreover, the study of the free theory is always a first step before considering interacting theories and the respective violating terms which is the case of non-Abelian theories as well as the Abelian theory with fermions.

\section{Pure Yang-Mills theory with Lorentz violation}\label{YM}

From now on, unless the contrary is said, we consider pure\footnote{Just like the Abelian case, we are not considering fermions.} Yang-Mills theory for the $SU(N)$ symmetry group with Lorentz violation. The gauge fields are algebra-valued $A_{\mu}=A^a_{\mu}T^a$, where $T^a$ are the generators of the $SU(N)$ algebra. They are chosen to be anti-Hermitian and have vanishing trace. The typical Lie algebra is given by $[T^a,T^b]=f^{abc}T^c$, where $f^{abc}$ are the skew-symmetric structure constants. The latin indices run as $\left\{a,b,c,\dots\right\}\;\in\;\left\{1,2,\dots,N^2-1\right\}$.

The model is described by the following action\footnote{No confusion is expected with the Abelian case.} \cite{Colladay:2006rk}
\begin{eqnarray}
\Sigma_0&=&S_{YM}+\Sigma_{LVE}+\Sigma_{LVO}\;,
\label{0}
\end{eqnarray}
where
\begin{eqnarray}
S_{YM}&=&\frac{1}{4}\int d^4xF^a_{\mu\nu}F^a_{\mu\nu}\;,
\label{1}
\end{eqnarray}
is the classical Yang-Mills action. The field strength is defined as $F^a_{\mu\nu}=\partial_{\mu}A^a_{\nu}-\partial_{\nu}A^a_{\mu}-gf^{abc}A^b_{\mu}A^c_{\nu}$. The Lorentz violating sector of even CPT is
\begin{eqnarray}
\Sigma_{LVE}&=&\frac{1}{4}\int d^4x\kappa_{\alpha\beta\mu\nu}F^a_{\alpha\beta}F^a_{\mu\nu}\;,
\label{2}
\end{eqnarray}
and odd CPT Lorentz violation term is
\begin{eqnarray}
\Sigma_{LVO}&=&\int d^4x\epsilon_{\mu\nu\alpha\beta}v_{\mu}\left(A^a_{\nu}\partial_{\alpha}A^a_{\beta}+\frac{g}{3}f^{abc}A^a_{\nu}A^b_{\alpha}A^c_{\beta}\right)\;.
\label{3A}
\end{eqnarray}
The Lorentz violation is characterized by the fields $v_{\mu}$, with mass dimension 1, and $\kappa_{\alpha\beta\mu\nu}$, which is dimensionless. These tensors have the same symmetry properties of those described in Sect.~\ref{QED} for the Abelian case.

\subsection{BRST quantization and the restoration of Lorentz symmetry} \label{QUATIZATION}

In the process of quantization of pure Yang-Mills theory with Lorentz violation, gauge fixing is also required. Inhere, we employ the BRST quantization method and adopt the Landau gauge condition $\partial_\mu A^a_\mu =0$. Thus, besides the gluon field, we also need the Lautrup-Nakanishi field $b^a$ and the Faddeev-Popov ghost and anti-ghost fields, namely, $c^a$ and $\bar{c}^a$, respectively. The BRST transformations of the fields are
\begin{eqnarray}
sA^a_{\mu}&=&-D^{ab}_\mu c^b\;,\nonumber\\
sc^a&=&\frac{g}{2}f^{abc}c^bc^c\;,\nonumber\\
s\bar{c}^a&=&b^a\;,\nonumber\\
sb^a&=&0\;,
\label{4a}
\end{eqnarray}
where $D^{ab}_{\mu}\;=\;\delta^{ab}\partial_\mu -gf^{abc}A^c_\mu$ is the covariant derivative. Thus, the Landau gauge fixed action is
\begin{eqnarray}
\Sigma_0&=&S_{YM}+\Sigma_{LVE}+\Sigma_{LVO}+\Sigma_{gf}\;,
\label{4b}
\end{eqnarray}
where 
\begin{eqnarray}
\Sigma_{gf}&=&s\int d^4x\bar{c}^a\partial_{\mu}A^a_{\mu}=\int d^4x\left(b^a\partial_{\mu}A^a_{\mu}+\bar{c}^a\partial_\mu D^{ab}_\mu c^b\right)\;,
\label{6}
\end{eqnarray}
is the gauge fixing action enforcing the Landau gauge condition. The Landau gauge is chosen here due to the same reasons of the Abelian case\footnote{Nevertheless, the renormalizability of YM theories with Lorentz violation could also be analyzed in other renormalizable gauges, e.g., the linear covariant $\xi$-gauges, the Maximal Abelian Gauge (MAG) and Curci-Ferrari gauge. All of them are very important in non-perturbative QCD studies. However, in the last two cases, they consist in non-linear gauges, a fact that demands the introduction of quartic ghost interacting terms for renormalizability and generate a large amount of extra counterterms, turning the whole analysis much less interesting and much more technical. The linear covariant gauges could be easily implemented, although extra terms depending on the gauge parameter would appear. However, as mentioned above, the Landau gauge is a natural fixed point of the linear covariant gauges, making them equivalent in some level.}. The quantum numbers of the fields and background tensors are the same as in the Abelian case, see Table \ref{tableA1}.

To deal with the renormalizability issue we will proceed in the same way we did in the Sec.~\ref{QED}, namely, we replace each background tensor by an external source and, possibly, its BRST doublet counterpart. However, the non-Abelian case is a bit more subtle than the Abelian case. For instance, let us take the Chern-Simons term: To ensure the renormalizability of the model we need two BRST doublets, one coupled to the bilinear term and another to the trilinear term in the gauge field. Both terms have to be treated separately since they are independent composite operators (in the Abelian case the Chern-Simons term has only one composite operator), see \eqref{7} below. The set of sources are characterized by
\begin{eqnarray}
s\bar{\kappa}_{\alpha\beta\mu\nu}&=&0\;,\nonumber\\
s\lambda_{\mu\nu\alpha}&=&J_{\mu\nu\alpha}\;,\nonumber\\
sJ_{\mu\nu\alpha}&=&0\;,\nonumber\\
s\eta_{\mu\nu\alpha}&=&\tau_{\mu\nu\alpha}\;,\nonumber\\
s\tau_{\mu\nu\alpha}&=&0\;.
\label{9}
\end{eqnarray}
Eventually, these sources will attain the following physical values
\begin{eqnarray}
J_{\mu\nu\alpha}\mid_{phys}&=&\tau_{\mu\nu\alpha}\mid_{phys}\;=\;v_{\beta}\epsilon_{\beta\mu\nu\alpha}\;,\nonumber\\
\lambda_{\mu\nu\alpha}\mid_{phys}&=&\eta_{\mu\nu\alpha}\mid_{phys}\;=\;0\;,\nonumber\\
\bar{\kappa}_{\alpha\beta\mu\nu}\mid_{phys}&=&\kappa_{\alpha\beta\mu\nu}\;.
\label{12}
\end{eqnarray}
Thus, we replace the action \eqref{4b} by\footnote{Since the Lorentz breaking is controlled by the external sources, we rename the original actions without the letter "V", of violation.}
\begin{eqnarray}
\Sigma'&=&S_{YM}+\Sigma_{LO}+\Sigma_{LE}+\Sigma_{gf}\ ,
\label{5}
\end{eqnarray}
where, now,
\begin{eqnarray}
\Sigma_{LE}&=&\frac{1}{4}\int d^4x\bar{\kappa}_{\alpha\beta\mu\nu}F^a_{\alpha\beta}F^a_{\mu\nu}\;,\nonumber\\
\Sigma_{LO}&=&s\int d^4x\left(\lambda_{\mu\nu\alpha}A^a_{\mu}\partial_{\nu}A^a_{\alpha}+\frac{g}{3}\eta_{\mu\nu\alpha}f^{abc}A^a_{\mu}A^b_{\nu}A^c_{\alpha}\right)\;.\nonumber\\
&=&\int d^4x\left[J_{\mu\nu\alpha}A^a_{\mu}\partial_{\nu}A^a_{\alpha}+\frac{g}{3}\tau_{\mu\nu\alpha}f^{abc}A^a_{\mu}A^b_{\nu}A^c_{\alpha}+\lambda_{\mu\nu\alpha}\partial_{\mu}c^a\partial_{\nu}A^a_{\alpha}+\right.\nonumber\\
&+&\left.g(\eta_{\mu\nu\alpha}-\lambda_{\mu\nu\alpha})f^{abc}A^a_{\mu}A^b_{\nu}\partial_{\alpha}c^c\right]\;.
\label{7}
\end{eqnarray}
is the embedding of the Lorentz violating bosonic sector. It is a trivial exercise to check that the new action is BRST invariant. The quantum numbers of the auxiliary sources follow the quantum numbers of the background fields, as displayed in table \ref{table3}.
\begin{table}[h]
\centering
\begin{tabular}{|c|c|c|c|c|c|c|c|}
	\hline 
sources & $\Omega$ & $L$ & $\lambda$ & $J$ & $\eta$ & $\tau$ & $\bar{\kappa}$  \\
	\hline 
UV dimension & $3$ & $4$ & $1$ & $1$ &$1$ & $1$ & $0$ \\ 
Ghost number & $-1$ & $-2$ & $-1$ & $0$& $-1$ & $0$ & $0$ \\ 
\hline 
\end{tabular}
\caption{Quantum numbers of the sources.}
\label{table3}
\end{table}

To face the issue of the renormalizability of the model, we need one last set of external BRST invariant sources, namely, $\Omega$ and $L$, in order to control the non-linear BRST transformations of the original fields,
\begin{eqnarray}
\Sigma_{ext}&=&s\int d^4x\left(-\Omega^a_{\mu}A^a_{\mu}+L^ac^a\right)\;,\nonumber\\
&=&\int d^4x\left(-\Omega^a_{\mu}D^{ab}_{\mu}c^b+\frac{g}{2}f^{abc}L^ac^bc^c\right)\;
\label{12A}
\end{eqnarray}

Still, from power counting analysis, and from BRST symmetry, extra bilinear terms in the gauge fields coupled to the auxiliary sources are allowed to be added to the action, namely
\begin{eqnarray}
\Sigma_{LCO}&=&s\int d^4x\left\{\left(\alpha_1\lambda_{\mu\nu\alpha}J_{\mu\nu\alpha}+\alpha_2\lambda_{\mu\nu\alpha}\tau_{\mu\nu\alpha}+\alpha_3\eta_{\mu\nu\alpha}J_{\mu\nu\alpha}+\alpha_4\eta_{\mu\nu\alpha}\tau_{\mu\nu\alpha}\right)\frac{1}{2}A^a_\beta A^a_\beta
+\right.\nonumber\\
&+&\left.\left(\beta_1\lambda_{\mu\alpha\beta}J_{\nu\alpha\beta}+\beta_2\lambda_{\mu\alpha\beta}\tau_{\nu\alpha\beta}+\beta_3\eta_{\mu\alpha\beta}J_{\nu\alpha\beta}+\beta_4\eta_{\mu\alpha\beta}\tau_{\nu\alpha\beta}\right)A^a_\mu A^a_\nu
+\right.\nonumber\\
&+&\left.
\bar{\kappa}_{\alpha\beta\mu\nu}\left(\gamma_1\lambda_{\alpha\beta\rho}J_{\mu\nu\rho}+\gamma_2\lambda_{\alpha\beta\rho}\tau_{\mu\nu\rho}+\gamma_3\eta_{\alpha\beta\rho}J_{\mu\nu\rho}+\gamma_4\eta_{\alpha\beta\rho}\tau_{\mu\nu\rho}\right)\frac{1}{2}A^a_{\sigma}A^a_{\sigma}+\right.\nonumber\\
&+&\left.\bar{\kappa}_{\alpha\beta\mu\nu}\left(\chi_1\lambda_{\beta\rho\sigma}J_{\nu\rho\sigma}+\chi_2\lambda_{\beta\rho\sigma}\tau_{\nu\rho\sigma}+\chi_3\eta_{\beta\rho\sigma}J_{\nu\rho\sigma}+\chi_4\eta_{\beta\rho\sigma}\tau_{\nu\rho\sigma}\right)A^a_{\alpha}A^a_{\mu}
+\right.\nonumber\\
&+&\left.\bar{\kappa}_{\alpha\rho\sigma\delta}\left(\varrho_1\lambda_{\nu\rho\delta}J_{\mu\alpha\sigma}+\varrho_2\lambda_{\nu\rho\delta}\tau_{\mu\alpha\sigma}+\varrho_3\eta_{\nu\rho\delta}\tau_{\mu\alpha\sigma}+\varrho_4\eta_{\nu\rho\delta}\tau_{\mu\alpha\sigma}\right)A^a_{\mu}A^a_{\nu}\right\}\nonumber\\
&=&\int d^4x\left\{\left(\alpha_1J_{\mu\nu\alpha}J_{\mu\nu\alpha}+\alpha_2J_{\mu\nu\alpha}\tau_{\mu\nu\alpha}+\alpha_3\tau_{\mu\nu\alpha}J_{\mu\nu\alpha}+\alpha_4\tau_{\mu\nu\alpha}\tau_{\mu\nu\alpha}\right)\frac{1}{2}A^a_\beta A^a_\beta
+\right.\nonumber\\
&+&\left.\left(\beta_1J_{\mu\alpha\beta}J_{\nu\alpha\beta}+\beta_2J_{\mu\alpha\beta}\tau_{\nu\alpha\beta}+\beta_3\tau_{\mu\alpha\beta}J_{\nu\alpha\beta}+\beta_4\tau_{\mu\alpha\beta}\tau_{\nu\alpha\beta}\right)A^a_\mu A^a_\nu
+\right.\nonumber\\
&+&\left.
\bar{\kappa}_{\alpha\beta\mu\nu}\left(\gamma_1J_{\alpha\beta\rho}J_{\mu\nu\rho}+\gamma_2J_{\alpha\beta\rho}\tau_{\mu\nu\rho}+\gamma_3\tau_{\alpha\beta\rho}J_{\mu\nu\rho}+\gamma_4\tau_{\alpha\beta\rho}\tau_{\mu\nu\rho}\right)\frac{1}{2}A^a_{\sigma}A^a_{\sigma}+\right.\nonumber\\
&+&\left.\bar{\kappa}_{\alpha\beta\mu\nu}\left(\chi_1J_{\beta\rho\sigma}J_{\nu\rho\sigma}+\chi_2J_{\beta\rho\sigma}\tau_{\nu\rho\sigma}+\chi_3\tau_{\beta\rho\sigma}J_{\nu\rho\sigma}+\chi_4\tau_{\beta\rho\sigma}\tau_{\nu\rho\sigma}\right)A^a_{\alpha}A^a_{\mu}
+\right.\nonumber\\
&+&\left.\bar{\kappa}_{\alpha\rho\sigma\delta}\left(\varrho_1J_{\nu\rho\delta}J_{\mu\alpha\sigma}+\varrho_2J_{\nu\rho\delta}J_{\mu\alpha\sigma}+\varrho_3\tau_{\nu\rho\delta}J_{\mu\alpha\sigma}+\varrho_4\tau_{\nu\rho\delta}J_{\mu\alpha\sigma}\right)A^a_{\mu}A^a_{\nu}+
\right.\nonumber\\
&+&\left.
\left(\alpha_1\lambda_{\mu\nu\alpha}J_{\mu\nu\alpha}+\alpha_2\lambda_{\mu\nu\alpha}\tau_{\mu\nu\alpha}+\alpha_3\eta_{\mu\nu\alpha}J_{\mu\nu\alpha}+\alpha_4\eta_{\mu\nu\alpha}\tau_{\mu\nu\alpha}\right)A^a_\beta \partial_\beta c^a
+\right.\nonumber\\
&+&\left.\left(\beta_1\lambda_{\mu\alpha\beta}J_{\nu\alpha\beta}+\beta_2\lambda_{\mu\alpha\beta}\tau_{\nu\alpha\beta}+\beta_3\eta_{\mu\alpha\beta}J_{\nu\alpha\beta}+\beta_4\eta_{\mu\alpha\beta}\tau_{\nu\alpha\beta}\right)(A^a_\mu \partial_\nu c^a+\partial_\mu c^aA^a_\nu)
+\right.\nonumber\\
&+&\left.
\bar{\kappa}_{\alpha\beta\mu\nu}\left(\gamma_1\lambda_{\alpha\beta\rho}J_{\mu\nu\rho}+\gamma_2\lambda_{\alpha\beta\rho}\tau_{\mu\nu\rho}+\gamma_3\eta_{\alpha\beta\rho}J_{\mu\nu\rho}+\gamma_4\eta_{\alpha\beta\rho}\tau_{\mu\nu\rho}\right)A^a_{\sigma}\partial_{\sigma}c^a+\right.\nonumber\\
&+&\left.\bar{\kappa}_{\alpha\beta\mu\nu}\left(\chi_1\lambda_{\beta\rho\sigma}J_{\nu\rho\sigma}+\chi_2\lambda_{\beta\rho\sigma}\tau_{\nu\rho\sigma}+\chi_3\eta_{\beta\rho\sigma}J_{\nu\rho\sigma}+\chi_4\eta_{\beta\rho\sigma}\tau_{\nu\rho\sigma}\right)(A^a_\alpha \partial_\mu c^a+\partial_\alpha c^aA^a_\mu)
+\right.\nonumber\\
&+&\left.\bar{\kappa}_{\alpha\rho\sigma\delta}\left(\varrho_1\lambda_{\nu\rho\delta}J_{\mu\alpha\sigma}+\varrho_2\lambda_{\nu\rho\delta}\tau_{\mu\alpha\sigma}+\varrho_3\eta_{\nu\rho\delta}J_{\mu\alpha\sigma}+\varrho_4\eta_{\nu\rho\delta}\tau_{\mu\alpha\sigma}\right)(A^a_\mu \partial_\nu c^a+\partial_\mu c^aA^a_\nu)\right\}\;.\nonumber\\
\label{12A1}
\end{eqnarray}
Clearly, a term of this type does not arise in the Abelian model. This property is due to the fact that the Abelian ghost equation is a non integrated identity, making it stronger than its non-Abelian version (we will discuss this issue after we define the physical action \eqref{15A}). Just like the Abelian case, a vacuum action, \textit{i.e.}, a term that depends only on the sources, is also allowed
\begin{eqnarray}
\Sigma_V&=&s\int d^4x\left\{\zeta_1 \lambda_{\mu\nu\alpha}J_{\mu\beta\gamma}J_{\nu\beta\kappa}J_{\gamma\kappa\alpha}+\zeta_2 \lambda_{\mu\nu\alpha}J_{\mu\beta\gamma}J_{\nu\beta\kappa}\tau_{\gamma\kappa\alpha}+\zeta_3 \lambda_{\mu\nu\alpha}J_{\mu\beta\gamma}\tau_{\nu\beta\kappa}J_{\gamma\kappa\alpha}+\zeta_4 \lambda_{\mu\nu\alpha}J_{\mu\beta\gamma}\tau_{\nu\beta\kappa}\tau_{\gamma\kappa\alpha}
+\right.\nonumber\\
&+&\left.
\zeta_5 \lambda_{\mu\nu\alpha}\tau_{\mu\beta\gamma}J_{\nu\beta\kappa}J_{\gamma\kappa\alpha}+\zeta_6 \lambda_{\mu\nu\alpha}\tau_{\mu\beta\gamma}J_{\nu\beta\kappa}\tau_{\gamma\kappa\alpha}+\zeta_7 \lambda_{\mu\nu\alpha}\tau_{\mu\beta\gamma}\tau_{\nu\beta\kappa}J_{\gamma\kappa\alpha}+\zeta_8 \lambda_{\mu\nu\alpha}\tau_{\mu\beta\gamma}\tau_{\nu\beta\kappa}\tau_{\gamma\kappa\alpha}
+\right.\nonumber\\
&+&\left.
\zeta_9 \eta_{\mu\nu\alpha}J_{\mu\beta\gamma}J_{\nu\beta\kappa}J_{\gamma\kappa\alpha}+\zeta_{10} \eta_{\mu\nu\alpha}J_{\mu\beta\gamma}J_{\nu\beta\kappa}\tau_{\gamma\kappa\alpha}+\zeta_{11} \eta_{\mu\nu\alpha}J_{\mu\beta\gamma}\tau_{\nu\beta\kappa}J_{\gamma\kappa\alpha}+\zeta_{12} \eta_{\mu\nu\alpha}J_{\mu\beta\gamma}\tau_{\nu\beta\kappa}\tau_{\gamma\kappa\alpha}
+\right.\nonumber\\
&+&\left.
\zeta_{13} \eta_{\mu\nu\alpha}\tau_{\mu\beta\gamma}J_{\nu\beta\kappa}J_{\gamma\kappa\alpha}+\zeta_{14} \eta_{\mu\nu\alpha}\tau_{\mu\beta\gamma}J_{\nu\beta\kappa}\tau_{\gamma\kappa\alpha}+\zeta_{15} \eta_{\mu\nu\alpha}\tau_{\mu\beta\gamma}\tau_{\nu\beta\kappa}J_{\gamma\kappa\alpha}+\zeta_{16} \eta_{\mu\nu\alpha}\tau_{\mu\beta\gamma}\tau_{\nu\beta\kappa}\tau_{\gamma\kappa\alpha}
+\right.\nonumber\\
&+&\left.\bar{\kappa}_{\mu\nu\alpha\beta}\left(\vartheta_1\lambda_{\mu\rho\omega}J_{\nu\rho\sigma}J_{\alpha\omega\delta}J_{\beta\sigma\delta}+\vartheta_2\lambda_{\mu\rho\omega}J_{\nu\rho\sigma}J_{\alpha\omega\delta}\tau_{\beta\sigma\delta}+\vartheta_3\lambda_{\mu\rho\omega}J_{\nu\rho\sigma}\tau_{\alpha\omega\delta}J_{\beta\sigma\delta}+\vartheta_4\lambda_{\mu\rho\omega}J_{\nu\rho\sigma}\tau_{\alpha\omega\delta}\tau_{\beta\sigma\delta}+\right.\right.\nonumber\\
&+&\left.\left.\vartheta_5\lambda_{\mu\rho\omega}\tau_{\nu\rho\sigma}J_{\alpha\omega\delta}J_{\beta\sigma\delta}+\vartheta_6\lambda_{\mu\rho\omega}\tau_{\nu\rho\sigma}J_{\alpha\omega\delta}\tau_{\beta\sigma\delta}+\vartheta_7\lambda_{\mu\rho\omega}\tau_{\nu\rho\sigma}\tau_{\alpha\omega\delta}J_{\beta\sigma\delta}+\vartheta_8\lambda_{\mu\rho\omega}\tau_{\nu\rho\sigma}\tau_{\alpha\omega\delta}\tau_{\beta\sigma\delta}+\right.\right.\nonumber\\
&+&\left.\left.
\vartheta_9\eta_{\mu\rho\omega}J_{\nu\rho\sigma}J_{\alpha\omega\delta}J_{\beta\sigma\delta}+\vartheta_{10}\eta_{\mu\rho\omega}J_{\nu\rho\sigma}J_{\alpha\omega\delta}\tau_{\beta\sigma\delta}+\vartheta_{11}\eta_{\mu\rho\omega}J_{\nu\rho\sigma}\tau_{\alpha\omega\delta}J_{\beta\sigma\delta}+\vartheta_{12}\eta_{\mu\rho\omega}J_{\nu\rho\sigma}\tau_{\alpha\omega\delta}\tau_{\beta\sigma\delta}+\right.\right.\nonumber\\
&+&\left.\left.
\vartheta_{13}\eta_{\mu\rho\omega}\tau_{\nu\rho\sigma}J_{\alpha\omega\delta}J_{\beta\sigma\delta}+\vartheta_{14}\eta_{\mu\rho\omega}\tau_{\nu\rho\sigma}J_{\alpha\omega\delta}\tau_{\beta\sigma\delta}+\vartheta_{15}\eta_{\mu\rho\omega}\tau_{\nu\rho\sigma}\tau_{\alpha\omega\delta}J_{\beta\sigma\delta}+\vartheta_{16}\eta_{\mu\rho\omega}\tau_{\nu\rho\sigma}\tau_{\alpha\omega\delta}\tau_{\beta\sigma\delta}\right)\right\}\nonumber\\
&=&\int d^4x\left\{\zeta_1 J_{\mu\nu\alpha}J_{\mu\beta\gamma}J_{\nu\beta\kappa}J_{\gamma\kappa\alpha}+\zeta_2 J_{\mu\nu\alpha}J_{\mu\beta\gamma}J_{\nu\beta\kappa}\tau_{\gamma\kappa\alpha}+\zeta_3 J_{\mu\nu\alpha}J_{\mu\beta\gamma}\tau_{\nu\beta\kappa}J_{\gamma\kappa\alpha}+\zeta_4 J_{\mu\nu\alpha}J_{\mu\beta\gamma}\tau_{\nu\beta\kappa}\tau_{\gamma\kappa\alpha}
+\right.\nonumber\\
&+&\left.
\zeta_5 J_{\mu\nu\alpha}\tau_{\mu\beta\gamma}J_{\nu\beta\kappa}J_{\gamma\kappa\alpha}+\zeta_6 J_{\mu\nu\alpha}\tau_{\mu\beta\gamma}J_{\nu\beta\kappa}\tau_{\gamma\kappa\alpha}+\zeta_7 J_{\mu\nu\alpha}\tau_{\mu\beta\gamma}\tau_{\nu\beta\kappa}J_{\gamma\kappa\alpha}+\zeta_8 J_{\mu\nu\alpha}\tau_{\mu\beta\gamma}\tau_{\nu\beta\kappa}\tau_{\gamma\kappa\alpha}
+\right.\nonumber\\
&+&\left.
\zeta_9 \tau_{\mu\nu\alpha}J_{\mu\beta\gamma}J_{\nu\beta\kappa}J_{\gamma\kappa\alpha}+\zeta_{10} \tau_{\mu\nu\alpha}J_{\mu\beta\gamma}J_{\nu\beta\kappa}\tau_{\gamma\kappa\alpha}+\zeta_{11} \tau_{\mu\nu\alpha}J_{\mu\beta\gamma}\tau_{\nu\beta\kappa}J_{\gamma\kappa\alpha}+\zeta_{12} \tau_{\mu\nu\alpha}J_{\mu\beta\gamma}\tau_{\nu\beta\kappa}\tau_{\gamma\kappa\alpha}
+\right.\nonumber\\
&+&\left.
\zeta_{13} \tau_{\mu\nu\alpha}\tau_{\mu\beta\gamma}J_{\nu\beta\kappa}J_{\gamma\kappa\alpha}+\zeta_{14} \tau_{\mu\nu\alpha}\tau_{\mu\beta\gamma}J_{\nu\beta\kappa}\tau_{\gamma\kappa\alpha}+\zeta_{15} \tau_{\mu\nu\alpha}\tau_{\mu\beta\gamma}\tau_{\nu\beta\kappa}J_{\gamma\kappa\alpha}+\zeta_{16} \tau_{\mu\nu\alpha}\tau_{\mu\beta\gamma}\tau_{\nu\beta\kappa}\tau_{\gamma\kappa\alpha}
+\right.\nonumber\\
&+&\left.\bar{\kappa}_{\mu\nu\alpha\beta}\left(\vartheta_1J_{\mu\rho\omega}J_{\nu\rho\sigma}J_{\alpha\omega\delta}J_{\beta\sigma\delta}+\vartheta_2J_{\mu\rho\omega}J_{\nu\rho\sigma}J_{\alpha\omega\delta}\tau_{\beta\sigma\delta}+\vartheta_3J_{\mu\rho\omega}J_{\nu\rho\sigma}\tau_{\alpha\omega\delta}J_{\beta\sigma\delta}+\vartheta_4J_{\mu\rho\omega}J_{\nu\rho\sigma}\tau_{\alpha\omega\delta}\tau_{\beta\sigma\delta}+\right.\right.\nonumber\\
&+&\left.\left.\vartheta_5J_{\mu\rho\omega}\tau_{\nu\rho\sigma}J_{\alpha\omega\delta}J_{\beta\sigma\delta}+\vartheta_6J_{\mu\rho\omega}\tau_{\nu\rho\sigma}J_{\alpha\omega\delta}\tau_{\beta\sigma\delta}+\vartheta_7J_{\mu\rho\omega}\tau_{\nu\rho\sigma}\tau_{\alpha\omega\delta}J_{\beta\sigma\delta}+\vartheta_8J_{\mu\rho\omega}\tau_{\nu\rho\sigma}\tau_{\alpha\omega\delta}\tau_{\beta\sigma\delta}+\right.\right.\nonumber\\
&+&\left.\left.
\vartheta_9\tau_{\mu\rho\omega}J_{\nu\rho\sigma}J_{\alpha\omega\delta}J_{\beta\sigma\delta}+\vartheta_{10}\tau_{\mu\rho\omega}J_{\nu\rho\sigma}J_{\alpha\omega\delta}\tau_{\beta\sigma\delta}+\vartheta_{11}\tau_{\mu\rho\omega}J_{\nu\rho\sigma}\tau_{\alpha\omega\delta}J_{\beta\sigma\delta}+\vartheta_{12}\tau_{\mu\rho\omega}J_{\nu\rho\sigma}\tau_{\alpha\omega\delta}\tau_{\beta\sigma\delta}+\right.\right.\nonumber\\
&+&\left.\left.
\vartheta_{13}\tau_{\mu\rho\omega}\tau_{\nu\rho\sigma}J_{\alpha\omega\delta}J_{\beta\sigma\delta}+\vartheta_{14}\tau_{\mu\rho\omega}\tau_{\nu\rho\sigma}J_{\alpha\omega\delta}\tau_{\beta\sigma\delta}+\vartheta_{15}\tau_{\mu\rho\omega}\tau_{\nu\rho\sigma}\tau_{\alpha\omega\delta}J_{\beta\sigma\delta}+\vartheta_{16}\tau_{\mu\rho\omega}\tau_{\nu\rho\sigma}\tau_{\alpha\omega\delta}\tau_{\beta\sigma\delta}\right)\right\}\;.
\label{12A1A}
\end{eqnarray}
Nonetheless, this action is larger than the Abelian action due to the number of auxiliary sources and their quantum numbers (see table .~\ref{table3}). The dimensionless parameters $\alpha_i,\beta_i,\gamma_i,\chi_i,\varrho_i$, with $i=\left\{1,\dots,4\right\}$ and $\zeta_j$ and $\vartheta_j$ with $j=\left\{1,\dots,16\right\}$ are required in order to absorb possible vacuum divergences. This extra term is inevitable due to the quantum numbers of the sources and the symmetries of the full action (see next Section). Moreover, some of the terms appearing in the actions \eqref{12A1} and \eqref{12A1A}, as we will see, always survive at the physical value of the sources. Thus, the vacuum of the model is directly affected. Just like the Abelian case, all infinite towers on the dimensionless source can be rearranged and redefined as the same original terms. The full action is then
\begin{eqnarray}
\Sigma&=&\Sigma'+\Sigma_{ext}+\Sigma_{LCO}+\Sigma_V\:.
\label{15}
\end{eqnarray}
At the physical value of the sources \eqref{12}, the action \eqref{15} reduces to
\begin{eqnarray}
\Sigma_{phys}&=&\frac{1}{4}\int d^4xF^a_{\mu\nu}F^a_{\mu\nu}+\frac{1}{4}\int d^4x\kappa_{\alpha\beta\mu\nu}F^a_{\alpha\beta}F^a_{\mu\nu}+\int d^4xv_{\beta}\epsilon_{\beta\mu\nu\alpha}\left(A^a_{\mu}\partial_{\nu}A^a_{\alpha}+\frac{g}{3}f^{abc}A^a_{\mu}A^b_{\nu}A^c_{\alpha}\right)+\nonumber\\
&+&\int d^4x\left(b^a\partial_{\mu}A^a_{\mu}+\bar{c}^a\partial_\mu D^{ab}_\mu c^b\right)+\int d^4x\left\{\left((3\alpha+2\beta)v^2-2(\gamma+\varrho)\kappa_{\alpha\sigma\rho\sigma}v_{\alpha}v_{\rho}\right) A^a_\mu A^a_\mu+\right.\nonumber\\
&-&\left.2\left(\beta v_\mu v_\nu +(\chi-\varrho)\kappa_{\sigma\mu\beta\nu}v_{\sigma}v_{\beta}-(\chi-\varrho)\kappa_{\mu\alpha\nu\alpha}v^2-2\varrho\kappa_{\rho\alpha\nu\alpha}v_{\rho} v_{\mu}\right)A^a_\mu A^a_\nu+6\zeta v^4+\right.\nonumber\\
&-&\left.2\vartheta\kappa_{\alpha\mu\sigma\mu}v_{\alpha}v_{\sigma}v^2\right\}\:,
\label{15A}
\end{eqnarray}
where
\begin{equation}
\alpha=\sum_{i=1}^{4}\alpha_i\;,\;\;\beta\;=\;\sum_{i=1}^{4}\beta_i\;,\;\;\chi\;=\;\sum_{i=1}^{4}\chi_i\;,\;\;\gamma\;=\;\sum_{i=1}^{4}\gamma_i\;,\;\;\varrho\;=\;\sum_{i=1}^{4}\varrho_i\;,\;\;\zeta=\sum_{j=1}^{16}\zeta_j\;,\;\;\vartheta\;=\;\sum_{j=1}^{16}\vartheta_j\;,
\label{15A0}
\end{equation}
from where it is evident that the vacuum is modified by the last two terms. Moreover, a typical Proca term is also generated, as well as a quadratic gauge field term with mixed indices. These two quadratic terms will change the tree-level propagator in a more dramatic way than the usual Lorentz violating Yang-Mills models. It is worth mention here that, in contrast to the Abelian case, the physical content of the gauge field will change drastically when the physical limit of the sources is taken. The only similarity with Abelian case is the emergence of a vacuum term. More specifically, by deforming the theory into a larger one and contracting it down back, the theory returns with extra terms (massive terms) that were not present before. We interpret this as a kind of mass (parameter) generation. Then, the field equations are indeed affected. This can also be seen from the propagators (see below), which are different from the typical non-Abelian Lorentz violating theories. The second point is that the pure source term $\Sigma_V$ also generates extra terms at the physical limit. These terms are constants and have no dependence on the quantum fields. They are pure vacuum terms, \emph{i.e.}, they do not affect the field equations but affect the vacuum of the theory. 

It is important to emphasize once again the fact that the mass terms not necessarily define a mass \emph{per se}. We refer to these terms as ``mass terms'' only because they appear as typical terms of massive theories. However, to determine if those masses are actually physical poles of the model is a task that goes beyond the scope of this work. Strictly speaking, those terms are related to mass parameters and not actual masses of the physical spectrum. In other words, if these mass parameters correspond, or not, to the propagation of massive physical modes, i.e., whether they are not tachyons nor ghosts. In QCD is quite typical the appearance of plenty mass parameters, however, they do not necessarily describe physical poles of the gluonic field, see for instance \cite{Dudal:2005na,Dudal:2011gd}. Anyhow, we have and will refer to these terms as mass terms. No confusion been expected from the reader.

For the propagator at the Landau gauge, a straightforward computation leads to (again, for technical reasons, we set $\kappa_{\alpha\beta\mu\nu}=0$)
\begin{eqnarray}
\langle A^a_{\mu}(k)A^b_{\nu}(-k)\rangle&=&\delta^{ab}\left(A\theta_{\mu\nu}+B\omega_{\mu\nu}+CS_{\mu\nu}+D\Sigma_{\mu\nu}+E\Lambda_{\mu\nu}\right)\;,
\label{15A1}
\end{eqnarray}
where
\begin{eqnarray}
A&=&\frac{k^2+\Delta v^2}{(k^2+\Delta v^2)^2-4[v^2k^2-(v_\alpha k_\alpha)^2]}\;,\nonumber\\
B&=&-(v_\alpha k_\alpha)D\;,\nonumber\\
C&=&\frac{2}{(k^2+\Delta v^2)^2-4[v^2k^2-(v_\alpha k_\alpha)^2]}\;,\nonumber\\
D&=&\frac{(v_\alpha k_\alpha)[\Omega(k^2+\Delta v^2)+4k^2]}{[k^2(k^2+\Delta v^2+\Omega v^2)-\Omega (v_\alpha k_\alpha)^2][(k^2+\Delta v^2)^2-4(v^2k^2-(v_\alpha k_\alpha)^2)]}\;,\nonumber\\
E&=&-\frac{k^2\left[\Omega(k^2+\Delta v^2)+4k^2\right]}{\left[k^2(k^2+\Delta v^2+ \Omega v^2)-\Omega(v_\alpha k_\alpha)^2\right]\left[(k^2+\Delta v^2)^2-4(v^2k^2-(v_\alpha k_\alpha)^2)\right]}\;,
\label{15A2}
\end{eqnarray}
and $\Delta=6\alpha+4\beta$ and $\Omega=-4\beta$. 

\section{Renormalizability}\label{RENORMALIZABILITY}

\subsection{Ward identities}

In order to prove the renormalizability of the model, we start by displaying the full set of Ward identities enjoyed by the action \eqref{15}.
\begin{itemize}
	\item Slavnov-Taylor identity
	\begin{eqnarray}
\mathcal{S}(\Sigma)&=&\int d^4x\left(\frac{\delta \Sigma}{\delta \Omega^a_{\mu}}\frac{\delta \Sigma}{\delta A^a_{\mu}}+\frac{\delta \Sigma}{\delta L^a}\frac{\delta \Sigma}{\delta c^a}+b^a\frac{\delta \Sigma}{\delta \bar{c}^a}+J_{\mu\nu\alpha}\frac{\delta \Sigma}{\delta\lambda_{\mu\nu\alpha}}+\tau_{\mu\nu\alpha}\frac{\delta \Sigma}{\delta\eta_{\mu\nu\alpha}}\right)=0\;.
\label{16}
\end{eqnarray}

\item Gauge fixing  equation and anti-ghost equation
\begin{eqnarray}
\frac{\delta \Sigma}{\delta b^a}&=&\partial_{\mu}A^a_{\mu}\;,\nonumber\\
\frac{\delta \Sigma}{\delta \bar{c}^a}+\partial_\mu \frac{\delta \Sigma}{\delta \Omega^a_\mu }&=&0\;.
\label{19}
\end{eqnarray}

\item Ghost equation
\begin{eqnarray}
\mathcal{G}^a\Sigma&=&\Delta^a_{cl}\;,
\label{20}
\end{eqnarray}
with 
\begin{eqnarray}
\mathcal{G}^a&=&\int d^4x\left(\frac{\delta}{\delta c^a}+gf^{abc}\bar{c}^b\frac{\delta}{\delta b^c}\right)\;,
\label{20A}
\end{eqnarray}
and
\begin{eqnarray}
\Delta^a_{cl}&=&\int d^4xgf^{abc}\left(\Omega^b_{\mu}A^c_{\mu}-L^bc^c\right)\;.
\label{20B}
\end{eqnarray}
\end{itemize}
At \eqref{19} and \eqref{20}, the breaking terms are linear in the fields. Thus, they will remain at classical level \cite{Piguet:1995er}.

\subsection{Most general counterterm}

In order to obtain the most general counterterm which can be freely added to the classical action $\Sigma$ at any order in perturbation theory, we define the most general local integrated polynomial $\Sigma^c$ with dimension bounded by four and vanishing ghost number. As usual, we impose the Ward identities (\ref{16}-\ref{20}) to the perturbed action $\Sigma+\varepsilon\Sigma^c$, where $\varepsilon$ is a small parameter. It is easy to find that the counterterm must obey the following constraints
\begin{eqnarray}
\mathcal{S}_{\Sigma}\Sigma^c&=&0\;,\nonumber\\
\frac{\delta \Sigma^c}{\delta b^a}&=&0\;,\nonumber\\
\left(\frac{\delta }{\delta \bar{c}^a}+\partial_\mu \frac{\delta }{\delta \Omega^a_\mu }\right)\Sigma^c&=&0\;,\nonumber\\
\mathcal{G}^a\Sigma^c&=&0\;,\label{22}
\end{eqnarray}
where the operator $\mathcal{S}_{\Sigma}$ is the nilpotent linearized Slavnov-Taylor operator,
\begin{eqnarray}
\mathcal{S}_{\Sigma}=\int d^4x\left(\frac{\delta \Sigma}{\delta \Omega^a_{\mu}}\frac{\delta}{\delta A^a_{\mu}}+\frac{\delta \Sigma}{\delta A^a_{\mu}}\frac{\delta }{\delta \Omega^a_{\mu}}+\frac{\delta \Sigma}{\delta L^a}\frac{\delta }{\delta c^a}+\frac{\delta \Sigma}{\delta  c^a}\frac{\delta}{\delta L^a}+b^a\frac{\delta}{\delta \bar{c}^a}+J_{\mu\nu\alpha}\frac{\delta }{\delta\lambda_{\mu\nu\alpha}}+\tau_{\mu\nu\alpha}\frac{\delta }{\delta\eta_{\mu\nu\alpha}}\right)\;.
\label{26}
\end{eqnarray}
The first constraint  of \eqref{22}, identifies the invariant counterterm as the solution of the cohomology problem for the operator $\mathcal{S}_{\Sigma}$ in the space of the integrated local field polynomials of dimension four and vanishing ghost number \cite{Piguet:1995er}. It follows that $\Sigma^c$ can be written as
\begin{eqnarray}
\Sigma^c&=&\frac{1}{4}\int d^4x\;a_0F^a_{\mu\nu}F^a_{\mu\nu}+\frac{1}{4}\int d^4x\;a_1\overline{\kappa}_{\alpha\beta\mu\nu}F^a_{\alpha\beta}F^a_{\mu\nu}+\mathcal{S}_{\Sigma}\Delta^{(-1)}\ ,
\label{27}
\end{eqnarray}
where $\Delta^{(-1)}$ is the most general local polynomial counterterm with dimension bounded by four and ghost number $-1$, given by\footnote{Just like the Abelian case, any infinite series in $\bar{\kappa}_{\mu\nu\alpha\beta}$ can be redefined as a single term linear in $\bar{\kappa}_{\mu\nu\alpha\beta}$.}
\begin{eqnarray}
\Delta^{(-1)}&=&\int d^4x\left\{a_2\Omega^a_\mu A^a_\mu +a_3\partial_\mu \bar{c}^aA^a_\mu +a_4L^ac^a+a_5\frac{1}{2}\bar{c}^ab^a+a_6\frac{g}{2}f^{abc}\bar{c}^a\bar{c}^bc^c+\right.\nonumber\\
&+&\left.\left(a_7\lambda_{\mu\nu\alpha}+a_8\eta_{\mu\nu\alpha}\right)A^a_{\mu}\partial_{\nu}A^a_{\alpha}+\frac{g}{3}\left(a_9\lambda_{\mu\nu\alpha}+a_{10}\eta_{\mu\nu\alpha}\right)f^{abc}A^a_{\mu}A^b_{\nu}A^c_{\alpha}+\right.\nonumber\\
&+&\left.\left(a_{11}\alpha_1\lambda_{\mu\nu\alpha}J_{\mu\nu\alpha}+a_{12}\alpha_2\lambda_{\mu\nu\alpha}\tau_{\mu\nu\alpha}+a_{13}\alpha_3\eta_{\mu\nu\alpha}J_{\mu\nu\alpha}+a_{14}\alpha_4\eta_{\mu\nu\alpha}\tau_{\mu\nu\alpha}\right)\frac{1}{2}A^a_\beta A^a_\beta
+\right.\nonumber\\
&+&\left.\left(a_{15}\beta_1\lambda_{\mu\alpha\beta}J_{\nu\alpha\beta}+a_{16}\beta_2\lambda_{\mu\alpha\beta}\tau_{\nu\alpha\beta}+a_{17}\beta_3\eta_{\mu\alpha\beta}J_{\nu\alpha\beta}+a_{18}\beta_4\eta_{\mu\alpha\beta}\tau_{\nu\alpha\beta}\right)A^a_\mu A^a_\nu
+\right.\nonumber\\
&+&\left.
\bar{\kappa}_{\alpha\beta\mu\nu}\left(a_{19}\gamma_1\lambda_{\alpha\beta\rho}J_{\mu\nu\rho}+a_{20}\gamma_2\lambda_{\alpha\beta\rho}\tau_{\mu\nu\rho}+a_{21}\gamma_3\eta_{\alpha\beta\rho}J_{\mu\nu\rho}+a_{22}\gamma_4\eta_{\alpha\beta\rho}\tau_{\mu\nu\rho}\right)\frac{1}{2}A^a_{\sigma}A^a_{\sigma}+\right.\nonumber\\
&+&\left.\bar{\kappa}_{\alpha\beta\mu\nu}\left(a_{23}\chi_1\lambda_{\beta\rho\sigma}J_{\nu\rho\sigma}+a_{24}\chi_2\lambda_{\beta\rho\sigma}\tau_{\nu\rho\sigma}+a_{25}\chi_3\eta_{\beta\rho\sigma}J_{\nu\rho\sigma}+a_{26}\chi_4\eta_{\beta\rho\sigma}\tau_{\nu\rho\sigma}\right)A^a_{\alpha}A^a_{\mu}
+\right.\nonumber\\
&+&\left.\bar{\kappa}_{\alpha\rho\sigma\delta}\left(a_{27}\varrho_1\lambda_{\nu\rho\delta}J_{\mu\alpha\sigma}+a_{28}\varrho_2\lambda_{\nu\rho\delta}\tau_{\mu\alpha\sigma}+a_{29}\varrho_3\eta_{\nu\rho\delta}\tau_{\mu\alpha\sigma}+a_{30}\varrho_4\eta_{\nu\rho\delta}\tau_{\mu\alpha\sigma}\right)A^a_{\mu}A^a_{\nu}+\right.\nonumber\\
&+&\left.a_{31}\zeta_1 \lambda_{\mu\nu\alpha}J_{\mu\beta\gamma}J_{\nu\beta\kappa}J_{\gamma\kappa\alpha}+a_{32}\zeta_2 \lambda_{\mu\nu\alpha}J_{\mu\beta\gamma}J_{\nu\beta\kappa}\tau_{\gamma\kappa\alpha}+a_{33}\zeta_3 \lambda_{\mu\nu\alpha}J_{\mu\beta\gamma}\tau_{\nu\beta\kappa}J_{\gamma\kappa\alpha}+\right.\nonumber\\
&+&\left.a_{34}\zeta_4 \lambda_{\mu\nu\alpha}J_{\mu\beta\gamma}\tau_{\nu\beta\kappa}\tau_{\gamma\kappa\alpha}
+
a_{35}\zeta_5 \lambda_{\mu\nu\alpha}\tau_{\mu\beta\gamma}J_{\nu\beta\kappa}J_{\gamma\kappa\alpha}+a_{36}\zeta_6 \lambda_{\mu\nu\alpha}\tau_{\mu\beta\gamma}J_{\nu\beta\kappa}\tau_{\gamma\kappa\alpha}+\right.\nonumber\\
&+&\left.a_{37}\zeta_7 \lambda_{\mu\nu\alpha}\tau_{\mu\beta\gamma}\tau_{\nu\beta\kappa}J_{\gamma\kappa\alpha}+a_{38}\zeta_8 \lambda_{\mu\nu\alpha}\tau_{\mu\beta\gamma}\tau_{\nu\beta\kappa}\tau_{\gamma\kappa\alpha}
+a_{39}\zeta_9 \eta_{\mu\nu\alpha}J_{\mu\beta\gamma}J_{\nu\beta\kappa}J_{\gamma\kappa\alpha}+\right.\nonumber\\
&+&\left.a_{40}\zeta_{10} \eta_{\mu\nu\alpha}J_{\mu\beta\gamma}J_{\nu\beta\kappa}\tau_{\gamma\kappa\alpha}+a_{41}\zeta_{11} \eta_{\mu\nu\alpha}J_{\mu\beta\gamma}\tau_{\nu\beta\kappa}J_{\gamma\kappa\alpha}+a_{42}\zeta_{12} \eta_{\mu\nu\alpha}J_{\mu\beta\gamma}\tau_{\nu\beta\kappa}\tau_{\gamma\kappa\alpha}
+\right.\nonumber\\
&+&\left.
a_{43}\zeta_{13} \eta_{\mu\nu\alpha}\tau_{\mu\beta\gamma}J_{\nu\beta\kappa}J_{\gamma\kappa\alpha}+a_{44}\zeta_{14} \eta_{\mu\nu\alpha}\tau_{\mu\beta\gamma}J_{\nu\beta\kappa}\tau_{\gamma\kappa\alpha}+a_{45}\zeta_{15} \eta_{\mu\nu\alpha}\tau_{\mu\beta\gamma}\tau_{\nu\beta\kappa}J_{\gamma\kappa\alpha}+\right.\nonumber\\
&+&\left.a_{46}\zeta_{16} \eta_{\mu\nu\alpha}\tau_{\mu\beta\gamma}\tau_{\nu\beta\kappa}\tau_{\gamma\kappa\alpha}
+\bar{\kappa}_{\mu\nu\alpha\beta}\left(a_{47}\vartheta_1\lambda_{\mu\rho\omega}J_{\nu\rho\sigma}J_{\alpha\omega\delta}J_{\beta\sigma\delta}+a_{48}\vartheta_2\lambda_{\mu\rho\omega}J_{\nu\rho\sigma}J_{\alpha\omega\delta}\tau_{\beta\sigma\delta}+\right.\right.\nonumber\\
&+&\left.\left.
a_{49}\vartheta_3\lambda_{\mu\rho\omega}J_{\nu\rho\sigma}\tau_{\alpha\omega\delta}J_{\beta\sigma\delta}+a_{50}\vartheta_4\lambda_{\mu\rho\omega}J_{\nu\rho\sigma}\tau_{\alpha\omega\delta}\tau_{\beta\sigma\delta}+
a_{51}\vartheta_5\lambda_{\mu\rho\omega}\tau_{\nu\rho\sigma}J_{\alpha\omega\delta}J_{\beta\sigma\delta}+\right.\right.\nonumber\\
&+&\left.\left.
a_{52}\vartheta_6\lambda_{\mu\rho\omega}\tau_{\nu\rho\sigma}J_{\alpha\omega\delta}\tau_{\beta\sigma\delta}+a_{53}\vartheta_7\lambda_{\mu\rho\omega}\tau_{\nu\rho\sigma}\tau_{\alpha\omega\delta}J_{\beta\sigma\delta}+a_{54}\vartheta_8\lambda_{\mu\rho\omega}\tau_{\nu\rho\sigma}\tau_{\alpha\omega\delta}\tau_{\beta\sigma\delta}+
+\right.\right.\nonumber\\
&+&\left.\left.a_{55}
\vartheta_9\eta_{\mu\rho\omega}J_{\nu\rho\sigma}J_{\alpha\omega\delta}J_{\beta\sigma\delta}+a_{56}\vartheta_{10}\eta_{\mu\rho\omega}J_{\nu\rho\sigma}J_{\alpha\omega\delta}\tau_{\beta\sigma\delta}+a_{57}\vartheta_{11}\eta_{\mu\rho\omega}J_{\nu\rho\sigma}\tau_{\alpha\omega\delta}J_{\beta\sigma\delta}+\right.\right.\nonumber\\
&+&\left.\left.a_{58}\vartheta_{12}\eta_{\mu\rho\omega}J_{\nu\rho\sigma}\tau_{\alpha\omega\delta}\tau_{\beta\sigma\delta}+
a_{59}\vartheta_{13}\eta_{\mu\rho\omega}\tau_{\nu\rho\sigma}J_{\alpha\omega\delta}J_{\beta\sigma\delta}+a_{60}\vartheta_{14}\eta_{\mu\rho\omega}\tau_{\nu\rho\sigma}J_{\alpha\omega\delta}\tau_{\beta\sigma\delta}+\right.\right.\nonumber\\
&+&\left.\left.a_{61}\vartheta_{15}\eta_{\mu\rho\omega}\tau_{\nu\rho\sigma}\tau_{\alpha\omega\delta}J_{\beta\sigma\delta}+a_{62}\vartheta_{16}\eta_{\mu\rho\omega}\tau_{\nu\rho\sigma}\tau_{\alpha\omega\delta}\tau_{\beta\sigma\delta}\right)\right\}\;,
\label{28}
\end{eqnarray}
where the parameters $a_i$  are free coefficients. The ghost equation implies $a_4=0$. Moreover, from the second or third equations in \eqref{22}, it follows that $a_2=a_3$. Still, from the second equation in \eqref{22} one finds that $a_5=a_6=0$. Then, it is straightforward to verify that the explicit form of the most general counterterm allowed by the Ward identities is the one given by \eqref{31} in the Appendix \ref{App1}.

\subsection{Stability}\label{STABILITY}

To finally prove the renormalizability of the the model we need to show that the counterterm $\Sigma^c$ can be reabsorbed by the original action $\Sigma$ by means of the redefinition of the fields, sources and parameters of the theory. Thus,
\begin{eqnarray}
\Sigma(\Phi, J, \xi)+\varepsilon\Sigma^c(\Phi, J, \xi)&=&\Sigma(\Phi_0, J_0, \xi_0)+\mathcal{O}(\varepsilon^2)\;,
\label{32}
\end{eqnarray}
where the bare fields, sources and parameters are defined as
\begin{eqnarray}
\Phi_0&=&Z^{1/2}_{\Phi}\Phi\:,\;\;\;\;\Phi \in \left\{A, b,\bar{c}, c\right\}\;,\nonumber\\
J_0&=&Z_JJ\;,\;\;\;\;\;\;\;\;\;J \in \left\{J,\lambda, \tau, \eta, \bar{\kappa}, \Omega, L\right\}\;, \nonumber\\
\xi_0&=&Z_{\xi}\xi\;,\;\;\;\;\;\;\;\;\;\xi \in \left\{g, \alpha_i,\beta_i,\chi_i,\gamma_i,\varrho_i,\zeta_j, \vartheta_j\right\}\;.
\label{33}
\end{eqnarray}
It is not difficult to check that this can be performed, proving the theory to be renormalizable to all orders in perturbation theory. Explicitly, the renormalization factors are listed bellow. 

For the independent renormalization factors of the gluon and coupling parameter, we have
\begin{eqnarray}
Z_A^{1/2}&=&1+\varepsilon \left(\frac{a_0}{2}+a_2\right)\;,\nonumber\\
Z_g&=&1-\varepsilon\frac{a_0}{2} \;,
\label{ren1}
\end{eqnarray}
while the renormalization factors of the ghosts, the Lautrup-Nakanishi field, $\Omega$ and $L$ sources are not independent:
\begin{eqnarray}
Z_c&=&Z_{\bar{c}}\;\;=\;\;Z_A^{-1/2}Z_g^{-1}\;,\nonumber\\
Z_{\Omega}&=&Z_A^{-1/4}Z_g^{-1/2}\;,\nonumber\\
Z_L&=&Z_b^{-1/2}\;=\;Z_A^{1/2}\;.
\label{ren2}
\end{eqnarray}
Thus, the renormalization of the standard Yang-Mills sector remains unchanged. For the sector associated with the vector $v_\mu$, \emph{i.e.}, the odd CPT breaking term, due to the quantum numbers of $J_{\mu\nu\alpha}$ and $\tau_{\mu\nu\alpha}$, there is a mixing between their respective operators, \emph{i.e.}, $A^a_{\mu}\partial_{\nu}A^a_{\alpha}$ and $gf^{abc}A^a_{\mu}A^b_{\nu}A^c_{\alpha}$. Thus, matrix renormalization is required, namely
\begin{equation}
\mathcal{J}_0=\mathcal{Z}_{\mathcal{J}}\mathcal{J}\;,\label{j}
\end{equation}
where $\mathcal{J}$ is a column matrix of sources that share the same quantum numbers. The quantity $Z_{\mathcal{J}}$ is a square matrix with the associated renormalization factors. In this case,
\begin{eqnarray}
\mathcal{J}_1=\begin{pmatrix}
J_{\mu\nu\alpha}\\
\tau_{\mu\nu\alpha}
\end{pmatrix}\;&\mathrm{and}&\;\mathcal{Z}_1=\begin{pmatrix}
Z_{JJ}& Z_{J\tau}\\
Z_{\tau J}& Z_{\tau\tau}
\end{pmatrix}\;\;=\;\;\mathbb{1}+\varepsilon\mathbb{A}\;,
\label{renm1}
\end{eqnarray}
where $\mathbb{A}$ is a matrix depending on $a_i$.
It is found that
\begin{equation}
\mathcal{Z}_1=\mathbb{1}+\varepsilon\begin{pmatrix}
a_7-a_0& a_8\\
a_9&a_{10}-a_0
\end{pmatrix}\;.
\end{equation}
The same rule will be used for the sources $\lambda_{\mu\nu\alpha}$ and $\eta_{\mu\nu\alpha}$, namely
\begin{eqnarray}
\mathcal{J}_2=\begin{pmatrix}
\lambda_{\mu\nu\alpha}\\
\eta_{\mu\nu\alpha}
\end{pmatrix}\;&\mathrm{and}&\;\mathcal{Z}_2=\begin{pmatrix}
Z_{\lambda\lambda}& Z_{\lambda\eta}\\
Z_{\eta\lambda}& Z_{\eta\eta}
\end{pmatrix}\;\;=\;\;\mathbb{1}+\varepsilon\mathbb{A}\;,
\label{renm1x}
\end{eqnarray}
where we find
\begin{equation}
\mathcal{Z}_2=\mathbb{1}+\varepsilon\begin{pmatrix}
\frac{a_2}{2}-\frac{a_0}{2}+a_7& a_8\\
a_9&\frac{a_2}{2}-\frac{a_0}{2}+a_{10}
\end{pmatrix}\;.
\end{equation}
For the even CPT breaking sector, the tensor $\kappa_{\mu\nu\alpha\beta}$ renormalizes through the factor
\begin{eqnarray}
Z_{\bar{\kappa}}&=&1+\varepsilon\left(a_1-a_0\right)\;,
\label{ren4}
\end{eqnarray}
while the renormalization factors of the corresponding parameters are given in the Appendix \ref{App2}. This ends the multiplicative renormalizability proof of the Lorentz violating pure Yang-Mills theory. An alternative, but equivalent, way to present the renormalization coefficients of the massless parameters is briefly displayed in the appendix \ref{Pr}.

\section{Conclusions}\label{FINAL}

In this work we have shown the multiplicative renormalizability of the Lorentz violating pure Yang-Mills theories, at least to all orders in perturbation theory. We have considered the Abelian and non-Abelian cases separately. In \cite{Colladay:2006rk}, through analytical renormalization technique, \textit{i.e.}, explicit 1-loop computation of the renormalization factors, the authors have already discussed the renormalizability of the non-Abelian case. In our prescription we employ only the algebraic technique \cite{Piguet:1995er}. The method allows for an all order analysis in perturbation theory. Remarkably, we have found that the odd CPT term induces mass terms for the non-Abelian gauge field while no mass is generated for the photon. It is known that massive parameters are already present due to the background $v_\mu$. However, the induced mass parameters come from the typical mass term on the action, namely $\nu^2A_\mu^aA_\mu^a$ and a mixing mass term $V_{\mu\nu}A_\mu^aA_\nu^a$, where $V_{\mu\nu}$ is a constant tensor (see \eqref{15A}). Furthermore, it was found that the renormalization properties of the usual sector of the these theories remain unaffected. The violating terms, however, have new renormalizations properties, except for the Abelian Chern-Simons-like term which does not renormalize.

In fact, at the Abelian case, there are only three new renormalizations, one is associated to the even sector of the breaking and the other two to pure vacuum terms. On the other hand, the odd sector of the Abelian breaking does not renormalize. At the non-Abelian case, however, fifty nine independent renormalizations are present. Besides the typical two renormalizations, the theory presents five independent renormalizations for the odd and even violating terms and thirty two parameters associated to a pure vacuum term. It is exactly the odd sector parameter which induces the extra mass terms which also renormalizes independently with twenty more parameters.

In \cite{Netto:2014faa}, the authors argue that quantum corrections in Lorentz and CPT violating QED in a curved manifold can induce, in a natural way, an effective action for gravity, besides this, as shown in \cite{Shapiro:2013bsa}, the original vacuum of the model is affected too. It is worth mention that, in the latter, the non-Abelian case is included. However, there exist some differences between the works \cite{Netto:2014faa,Shapiro:2013bsa} and the one presented here: the main one is that here we work in a flat-manifold, \textit{i.e.}, Euclidean space-time. Furthermore, besides the fact that the Lorentz violating coefficients have been treated here as local sources, their physical values are simply constant coefficients, in contrast with \cite{Netto:2014faa,Shapiro:2013bsa}. Moreover, in these works the even Lorentz violating CPT coefficient does not have double vanishing trace. A non vanishing double trace of the even Lorentz violating CPT coefficient could bring important consequences in a non-Abelian model, as for instance, the presence of dimension four operators \cite{Dudal:2008tg}, and also could bring consequences to the ghost sector of the model. A common assumption between our and \cite{deBerredoPeixoto:2006wz,Netto:2014faa,Shapiro:2013bsa} works was that higher towers in the dimensionless parameters (sources) are suppressed assuming their classical behaviour. In our case, however, nothing can be said about whether the vacuum terms presented here could bring cosmological effects, at least in phenomenological way, in contrast to what was discussed in \cite{Netto:2014faa,Shapiro:2013bsa}.

An interesting point to be studied is the explicit computation of the background tensors by applying the renormalization group equations combined with some extra condition for each for the tensors. For instance, following the Gribov-Zwanziger method, a minimal sensitivity principle could be applied. Such kind of condition may also be combined with phenomenological information in order to provide reliable bounds for these tensors. In this context, it will be important to choose a renormalization scheme which works in the present approach. The first reliable choice would be a minimal subtraction scheme because it works nicely in similar contexts such as the Gribov-Zwanziger analysis, see \cite{Dudal:2005na,Dudal:2011gd}.

Another interesting point would be the all orders proof of the renormalizability of the electroweak theory and QCD theory with Lorentz violation considering the fermionic and bosonic sectors \cite{Colladay:2009rb, Colladay:2007aj}. Moreover, the Gribov ambiguity problem \cite{Gribov:1977wm,Singer:1978dk,Sobreiro:2005ec,Pereira:2013aza} is also manifest at the Lorentz violating Yang-Mills action. Thus, the inclusion of the refined Gribov-Zwanziger terms could also lead to non-trivial effects that could be visualized through the propagators. In fact, the analysis of the poles of the propagators \eqref{B6A} and \eqref{15A1} and the respective restrictions on the backgrounds is currently under investigation \cite{WIP}. Nonetheless, all these analysis might be very difficult and tricky and, for this reason, we leave them for future investigation.

\appendix

\section{Counterterm}\label{App1}

The counterterm of the non-Abelian theory is found to be:
\begin{eqnarray}
\Sigma^c&=&a_0S_{YM}+a_1\Sigma_{LE}+a_2\int d^4x\left[\frac{\delta S_{YM}}{\delta A^a_\mu }A^a_\mu+\frac{\delta \Sigma_{LE}}{\delta A^a_\mu }A^a_\mu+\frac{\delta \Sigma_{LO}}{\delta A^a_\mu }A^a_\mu +\right.\nonumber\\
&+&\left.\left(\alpha_1J_{\mu\nu\alpha}J_{\mu\nu\alpha}+\alpha_2J_{\mu\nu\alpha}\tau_{\mu\nu\alpha}+\alpha_3\tau_{\mu\nu\alpha}J_{\mu\nu\alpha}+\alpha_4\tau_{\mu\nu\alpha}\tau_{\mu\nu\alpha}\right)A^a_\beta A^a_\beta
+\right.\nonumber\\
&+&\left.\left.2(\beta_1J_{\mu\alpha\beta}J_{\nu\alpha\beta}+\beta_2J_{\mu\alpha\beta}\tau_{\nu\alpha\beta}+\beta_3\tau_{\mu\alpha\beta}J_{\nu\alpha\beta}+\beta_4\tau_{\mu\alpha\beta}\tau_{\nu\alpha\beta}\right)A^a_\mu A^a_\nu
+\right.\nonumber\\
&+&\left.
\bar{\kappa}_{\alpha\beta\mu\nu}\left(\gamma_1J_{\alpha\beta\rho}J_{\mu\nu\rho}+\gamma_2J_{\alpha\beta\rho}\tau_{\mu\nu\rho}+\gamma_3\tau_{\alpha\beta\rho}J_{\mu\nu\rho}+\gamma_4\tau_{\alpha\beta\rho}\tau_{\mu\nu\rho}\right)A^a_{\sigma}A^a_{\sigma}+\right.\nonumber\\
&+&\left.2\bar{\kappa}_{\alpha\beta\mu\nu}\left(\chi_1J_{\beta\rho\sigma}J_{\nu\rho\sigma}+\chi_2J_{\beta\rho\sigma}\tau_{\nu\rho\sigma}+\chi_3\tau_{\beta\rho\sigma}J_{\nu\rho\sigma}+\chi_4\tau_{\beta\rho\sigma}\tau_{\nu\rho\sigma}\right)A^a_{\alpha}A^a_{\mu}
+\right.\nonumber\\
&+&\left.2\bar{\kappa}_{\alpha\rho\sigma\delta}\left(\varrho_1J_{\nu\rho\delta}J_{\mu\alpha\sigma}+\varrho_2J_{\nu\rho\delta}J_{\mu\alpha\sigma}+\varrho_3\tau_{\nu\rho\delta}J_{\mu\alpha\sigma}+\varrho_4\tau_{\nu\rho\delta}J_{\mu\alpha\sigma}\right)A^a_{\mu}A^a_{\nu}+
\right.\nonumber\\
&+&\left.
\left(\alpha_1\lambda_{\mu\nu\alpha}J_{\mu\nu\alpha}+\alpha_2\lambda_{\mu\nu\alpha}\tau_{\mu\nu\alpha}+\alpha_3\eta_{\mu\nu\alpha}J_{\mu\nu\alpha}+\alpha_4\eta_{\mu\nu\alpha}\tau_{\mu\nu\alpha}\right)A^a_\beta \partial_\beta c^a
+\right.\nonumber\\
&+&\left.\left(\beta_1\lambda_{\mu\alpha\beta}J_{\nu\alpha\beta}+\beta_2\lambda_{\mu\alpha\beta}\tau_{\nu\alpha\beta}+\beta_3\eta_{\mu\alpha\beta}J_{\nu\alpha\beta}+\beta_4\eta_{\mu\alpha\beta}\tau_{\nu\alpha\beta}\right)(A^a_\mu \partial_\nu c^a+\partial_\mu c^aA^a_\nu)
+\right.\nonumber\\
&+&\left.
\bar{\kappa}_{\alpha\beta\mu\nu}\left(\gamma_1\lambda_{\alpha\beta\rho}J_{\mu\nu\rho}+\gamma_2\lambda_{\alpha\beta\rho}\tau_{\mu\nu\rho}+\gamma_3\eta_{\alpha\beta\rho}J_{\mu\nu\rho}+\gamma_4\eta_{\alpha\beta\rho}\tau_{\mu\nu\rho}\right)A^a_{\sigma}\partial_{\sigma}c^a+\right.\nonumber\\
&+&\left.\bar{\kappa}_{\alpha\beta\mu\nu}\left(\chi_1\lambda_{\beta\rho\sigma}J_{\nu\rho\sigma}+\chi_2\lambda_{\beta\rho\sigma}\tau_{\nu\rho\sigma}+\chi_3\eta_{\beta\rho\sigma}J_{\nu\rho\sigma}+\chi_4\eta_{\beta\rho\sigma}\tau_{\nu\rho\sigma}\right)(A^a_\alpha \partial_\mu c^a+\partial_\alpha c^aA^a_\mu)
+\right.\nonumber\\
&+&\left.\bar{\kappa}_{\alpha\rho\sigma\delta}\left(\varrho_1\lambda_{\nu\rho\delta}J_{\mu\alpha\sigma}+\varrho_2\lambda_{\nu\rho\delta}\tau_{\mu\alpha\sigma}+\varrho_3\eta_{\nu\rho\delta}J_{\mu\alpha\sigma}+\varrho_4\eta_{\nu\rho\delta}\tau_{\mu\alpha\sigma}\right)(A^a_\mu \partial_\nu c^a+\partial_\mu c^aA^a_\nu)\right]+\nonumber\\
&+&\int d^4x\left\{J_{\mu\nu\alpha}\left(a_7A^a_{\mu}\partial_{\nu}A^a_{\alpha}+a_9\frac{g}{3}f^{abc}A^a_{\mu}A^b_{\nu}A^c_{\alpha}\right)+a_7\lambda_{\mu\nu\alpha}\partial_{\mu}c^a\partial_{\nu}A^a_{\alpha}+(a_9-a_7)g\lambda_{\mu\nu\alpha}f^{abc}A^a_\mu A^c_\alpha \partial_\nu c^b+\right.\nonumber\\
&+&\left.\tau_{\mu\nu\alpha}\left(a_8A^a_{\mu}\partial_{\nu}A^a_{\alpha}+a_{10}\frac{g}{3}f^{abc}A^a_{\mu}A^b_{\nu}A^c_{\alpha}\right)+a_8\eta_{\mu\nu\alpha}\partial_{\mu}c^a\partial_{\nu}A^a_{\alpha}+(a_{10}-a_8)g\eta_{\mu\nu\alpha}f^{abc}A^a_\mu A^c_\alpha \partial_\nu c^b+\right.\nonumber\\
&+&\left.\left(a_{11}\alpha_1J_{\mu\nu\alpha}J_{\mu\nu\alpha}+a_{12}\alpha_2J_{\mu\nu\alpha}\tau_{\mu\nu\alpha}+a_{13}\alpha_3\tau_{\mu\nu\alpha}J_{\mu\nu\alpha}+a_{14}\alpha_4\tau_{\mu\nu\alpha}\tau_{\mu\nu\alpha}\right)\frac{1}{2}A^a_\beta A^a_\beta
+\right.\nonumber\\
&+&\left.\left(a_{11}\alpha_1\lambda_{\mu\nu\alpha}J_{\mu\nu\alpha}+a_{12}\alpha_2\lambda_{\mu\nu\alpha}\tau_{\mu\nu\alpha}+a_{13}\alpha_3\eta_{\mu\nu\alpha}J_{\mu\nu\alpha}+a_{14}\alpha_4\eta_{\mu\nu\alpha}\tau_{\mu\nu\alpha}\right)A^a_\beta \partial_\beta
c^a+\right.\nonumber\\
&+&\left.\left(a_{15}\beta_1J_{\mu\alpha\beta}J_{\nu\alpha\beta}+a_{16}\beta_2J_{\mu\alpha\beta}\tau_{\nu\alpha\beta}+a_{17}\beta_3\tau_{\mu\alpha\beta}J_{\nu\alpha\beta}+a_{18}\beta_4\tau_{\mu\alpha\beta}\tau_{\nu\alpha\beta}\right)A^a_\mu A^a_\nu
+\right.\nonumber\\
&+&\left.\left(a_{15}\beta_1\lambda_{\mu\alpha\beta}J_{\nu\alpha\beta}+a_{16}\beta_2\lambda_{\mu\alpha\beta}\tau_{\nu\alpha\beta}+a_{17}\beta_3\eta_{\mu\alpha\beta}J_{\nu\alpha\beta}+a_{18}\beta_4\eta_{\mu\alpha\beta}\tau_{\nu\alpha\beta}\right)(A^a_\mu \partial_\nu c^a+\partial_\mu c^aA^a_\nu )
+\right.\nonumber\\
&+&\left.
\bar{\kappa}_{\alpha\beta\mu\nu}\left(a_{19}\gamma_1J_{\alpha\beta\rho}J_{\mu\nu\rho}+a_{20}\gamma_2J_{\alpha\beta\rho}\tau_{\mu\nu\rho}+a_{21}\gamma_3\tau_{\alpha\beta\rho}J_{\mu\nu\rho}+a_{22}\gamma_4\tau_{\alpha\beta\rho}\tau_{\mu\nu\rho}\right)\frac{1}{2}A^a_{\sigma}A^a_{\sigma}+\right.\nonumber\\
&+&\left.
\bar{\kappa}_{\alpha\beta\mu\nu}\left(a_{19}\gamma_1\lambda_{\alpha\beta\rho}J_{\mu\nu\rho}+a_{20}\gamma_2\lambda_{\alpha\beta\rho}\tau_{\mu\nu\rho}+a_{21}\gamma_3\eta_{\alpha\beta\rho}J_{\mu\nu\rho}+a_{22}\gamma_4\eta_{\alpha\beta\rho}\tau_{\mu\nu\rho}\right)A^a_{\sigma}\partial_{\sigma}c^a+\right.\nonumber\\
&+&\left.
\bar{\kappa}_{\alpha\beta\mu\nu}\left(a_{23}\chi_1J_{\beta\rho\sigma}J_{\nu\rho\sigma}+a_{24}\chi_2J_{\beta\rho\sigma}\tau_{\nu\rho\sigma}+a_{25}\chi_3\tau_{\beta\rho\sigma}J_{\nu\rho\sigma}+a_{26}\chi_4\tau_{\beta\rho\sigma}\tau_{\nu\rho\sigma}\right)A^a_{\alpha}A^a_{\mu}
+\right.\nonumber\\
&+&\left.
\bar{\kappa}_{\alpha\beta\mu\nu}\left(a_{23}\chi_1\lambda_{\beta\rho\sigma}J_{\nu\rho\sigma}+a_{24}\chi_2\lambda_{\beta\rho\sigma}\tau_{\nu\rho\sigma}+a_{25}\chi_3\eta_{\beta\rho\sigma}J_{\nu\rho\sigma}+a_{26}\chi_4\eta_{\beta\rho\sigma}\tau_{\nu\rho\sigma}\right)(A^a_\alpha \partial_\mu c^a+\partial_\mu c^aA^a_\alpha )
+\right.\nonumber\\
&+&\left.
\bar{\kappa}_{\alpha\rho\sigma\delta}\left(a_{27}\varrho_1J_{\nu\rho\delta}J_{\mu\alpha\sigma}+a_{28}\varrho_2J_{\nu\rho\delta}\tau_{\mu\alpha\sigma}+a_{29}\varrho_3\tau_{\nu\rho\delta}\tau_{\mu\alpha\sigma}+a_{30}\varrho_4\tau_{\nu\rho\delta}\tau_{\mu\alpha\sigma}\right)A^a_{\mu}A^a_{\nu}+\right.\nonumber\\
&+&\left.
\bar{\kappa}_{\alpha\rho\sigma\delta}\left(a_{27}\varrho_1\lambda_{\nu\rho\delta}J_{\mu\alpha\sigma}+a_{28}\varrho_2\lambda_{\nu\rho\delta}\tau_{\mu\alpha\sigma}+a_{29}\varrho_3\eta_{\nu\rho\delta}\tau_{\mu\alpha\sigma}+a_{30}\varrho_4\eta_{\nu\rho\delta}\tau_{\mu\alpha\sigma}\right)(A^a_\mu \partial_\nu c^a+\partial_\mu c^aA^a_\nu )+\right.\nonumber\\
&+&\left.
a_{31}\zeta_1 J_{\mu\nu\alpha}J_{\mu\beta\gamma}J_{\nu\beta\kappa}J_{\gamma\kappa\alpha}+a_{32}\zeta_2 J_{\mu\nu\alpha}J_{\mu\beta\gamma}J_{\nu\beta\kappa}\tau_{\gamma\kappa\alpha}+a_{33}\zeta_3 J_{\mu\nu\alpha}J_{\mu\beta\gamma}\tau_{\nu\beta\kappa}J_{\gamma\kappa\alpha}+\right.\nonumber\\
&+&\left.a_{34}\zeta_4 J_{\mu\nu\alpha}J_{\mu\beta\gamma}\tau_{\nu\beta\kappa}\tau_{\gamma\kappa\alpha}
+
a_{35}\zeta_5 J_{\mu\nu\alpha}\tau_{\mu\beta\gamma}J_{\nu\beta\kappa}J_{\gamma\kappa\alpha}+a_{36}\zeta_6 J_{\mu\nu\alpha}\tau_{\mu\beta\gamma}J_{\nu\beta\kappa}\tau_{\gamma\kappa\alpha}+\right.\nonumber\\
&+&\left.a_{37}\zeta_7 J_{\mu\nu\alpha}\tau_{\mu\beta\gamma}\tau_{\nu\beta\kappa}J_{\gamma\kappa\alpha}+a_{38}\zeta_8 J_{\mu\nu\alpha}\tau_{\mu\beta\gamma}\tau_{\nu\beta\kappa}\tau_{\gamma\kappa\alpha}
+a_{39}\zeta_9 \tau_{\mu\nu\alpha}J_{\mu\beta\gamma}J_{\nu\beta\kappa}J_{\gamma\kappa\alpha}+\right.\nonumber\\
&+&\left.a_{40}\zeta_{10} \tau_{\mu\nu\alpha}J_{\mu\beta\gamma}J_{\nu\beta\kappa}\tau_{\gamma\kappa\alpha}+a_{41}\zeta_{11} \tau_{\mu\nu\alpha}J_{\mu\beta\gamma}\tau_{\nu\beta\kappa}J_{\gamma\kappa\alpha}+a_{42}\zeta_{12} \tau_{\mu\nu\alpha}J_{\mu\beta\gamma}\tau_{\nu\beta\kappa}\tau_{\gamma\kappa\alpha}
+\right.\nonumber\\
&+&\left.
a_{43}\zeta_{13} \tau_{\mu\nu\alpha}\tau_{\mu\beta\gamma}J_{\nu\beta\kappa}J_{\gamma\kappa\alpha}+a_{44}\zeta_{14} \tau_{\mu\nu\alpha}\tau_{\mu\beta\gamma}J_{\nu\beta\kappa}\tau_{\gamma\kappa\alpha}+a_{45}\zeta_{15} \tau_{\mu\nu\alpha}\tau_{\mu\beta\gamma}\tau_{\nu\beta\kappa}J_{\gamma\kappa\alpha}+\right.\nonumber\\
&+&\left.a_{46}\zeta_{16} \tau_{\mu\nu\alpha}\tau_{\mu\beta\gamma}\tau_{\nu\beta\kappa}\tau_{\gamma\kappa\alpha}
+\bar{\kappa}_{\mu\nu\alpha\beta}\left(a_{47}\vartheta_1J_{\mu\rho\omega}J_{\nu\rho\sigma}J_{\alpha\omega\delta}J_{\beta\sigma\delta}+a_{48}\vartheta_2J_{\mu\rho\omega}J_{\nu\rho\sigma}J_{\alpha\omega\delta}\tau_{\beta\sigma\delta}+\right.\right.\nonumber\\
&+&\left.\left.
a_{49}\vartheta_3J_{\mu\rho\omega}J_{\nu\rho\sigma}\tau_{\alpha\omega\delta}J_{\beta\sigma\delta}+a_{50}\vartheta_4J_{\mu\rho\omega}J_{\nu\rho\sigma}\tau_{\alpha\omega\delta}\tau_{\beta\sigma\delta}+
a_{51}\vartheta_5J_{\mu\rho\omega}\tau_{\nu\rho\sigma}J_{\alpha\omega\delta}J_{\beta\sigma\delta}+\right.\right.\nonumber\\
&+&\left.\left.
a_{52}\vartheta_6J_{\mu\rho\omega}\tau_{\nu\rho\sigma}J_{\alpha\omega\delta}\tau_{\beta\sigma\delta}+a_{53}\vartheta_7J_{\mu\rho\omega}\tau_{\nu\rho\sigma}\tau_{\alpha\omega\delta}J_{\beta\sigma\delta}+a_{54}\vartheta_8J_{\mu\rho\omega}\tau_{\nu\rho\sigma}\tau_{\alpha\omega\delta}\tau_{\beta\sigma\delta}+\right.\right.\nonumber\\
&+&\left.\left.a_{55}
\vartheta_9\tau_{\mu\rho\omega}J_{\nu\rho\sigma}J_{\alpha\omega\delta}J_{\beta\sigma\delta}+a_{56}\vartheta_{10}\tau_{\mu\rho\omega}J_{\nu\rho\sigma}J_{\alpha\omega\delta}\tau_{\beta\sigma\delta}+a_{57}\vartheta_{11}\tau_{\mu\rho\omega}J_{\nu\rho\sigma}\tau_{\alpha\omega\delta}J_{\beta\sigma\delta}+\right.\right.\nonumber\\
&+&\left.\left.a_{58}\vartheta_{12}\tau_{\mu\rho\omega}J_{\nu\rho\sigma}\tau_{\alpha\omega\delta}\tau_{\beta\sigma\delta}+
a_{59}\vartheta_{13}\tau_{\mu\rho\omega}\tau_{\nu\rho\sigma}J_{\alpha\omega\delta}J_{\beta\sigma\delta}+a_{60}\vartheta_{14}\tau_{\mu\rho\omega}\tau_{\nu\rho\sigma}J_{\alpha\omega\delta}\tau_{\beta\sigma\delta}+\right.\right.\nonumber\\
&+&\left.\left.a_{61}\vartheta_{15}\tau_{\mu\rho\omega}\tau_{\nu\rho\sigma}\tau_{\alpha\omega\delta}J_{\beta\sigma\delta}+a_{62}\vartheta_{16}\tau_{\mu\rho\omega}\tau_{\nu\rho\sigma}\tau_{\alpha\omega\delta}\tau_{\beta\sigma\delta}\right)\right\}\;.\nonumber\\
\label{31}
\end{eqnarray}

\section{Renormalization factors of the parameters}\label{App2}

The renormalization factors of the dimensionless parameters are found to be:
\begin{align}
Z_{\alpha_1}&=1+\varepsilon\left(a_{11}-2a_7+a_0-\frac{\alpha_2+\alpha_3}{\alpha_1}a_9\right)\;,\displaybreak[3]\nonumber\\
Z_{\alpha_2}&=1+\varepsilon\left(a_{12}-a_7-a_{10}+a_0-\left(\frac{\alpha_1}{\alpha_2}a_8+\frac{\alpha_4}{\alpha_2}a_9\right)\right)\;,\displaybreak[3]\nonumber\\
Z_{\alpha_3}&=1+\varepsilon\left(a_{13}-a_7-a_{10}+a_0-\left(\frac{\alpha_1}{\alpha_3}a_8+\frac{\alpha_4}{\alpha_3}a_9\right)\right)\;,\displaybreak[3]\nonumber\\
Z_{\alpha_4}&=1+\varepsilon\left(a_{14}-2a_{10}+a_0-\frac{\alpha_2+\alpha_3}{\alpha_4}a_8\right)\;,\displaybreak[3]\nonumber\\
Z_{\beta_1}&=1+\varepsilon\left(a_{15}-2a_7+a_0-\frac{\beta_2+\beta_3}{\beta_1}a_9\right)\;,\displaybreak[3]\nonumber\\
Z_{\beta_2}&=1+\varepsilon\left(a_{16}-a_7-a_{10}+a_0-\left(\frac{\beta_1}{\beta_2}a_8+\frac{\beta_4}{\beta_2}a_9\right)\right)\;,\displaybreak[3]\nonumber\\
Z_{\beta_3}&=1+\varepsilon\left(a_{17}-a_7-a_{10}+a_0-\left(\frac{\beta_1}{\beta_3}a_8+\frac{\beta_4}{\beta_3}a_9\right)\right)\;,\displaybreak[3]\nonumber\\
Z_{\beta_4}&=1+\varepsilon\left(a_{18}-2a_{10}+a_0-\frac{\beta_2+\beta_3}{\beta_4}a_8\right)\;,\displaybreak[3]\nonumber\\
Z_{\gamma_1}&=1+\varepsilon\left(a_{19}-a_1-2a_7+2a_0-\frac{\gamma_2+\gamma_3}{\gamma_1}a_9\right)\;,\displaybreak[3]\nonumber\\
Z_{\gamma_2}&=1+\varepsilon\left(a_{20}-a_1-a_7-a_{10}+2a_0-\left(\frac{\gamma_1}{\gamma_2}a_8+\frac{\gamma_4}{\gamma_2}a_9\right)\right)\;,\displaybreak[3]\nonumber\\
Z_{\gamma_3}&=1+\varepsilon\left(a_{21}-a_1-a_7-a_{10}+2a_0-\left(\frac{\gamma_1}{\gamma_3}a_8+\frac{\gamma_4}{\gamma_3}a_9\right)\right)\;,\displaybreak[3]\nonumber\\
Z_{\gamma_4}&=1+\varepsilon\left(a_{22}-a_1-2a_{10}+2a_0-\frac{\gamma_2+\gamma_3}{\gamma_4}a_8\right)\;,\displaybreak[3]\nonumber\\
Z_{\chi_1}&=1+\varepsilon\left(a_{23}-a_1-2a_7+2a_0-\frac{\chi_2+\chi_3}{\chi_1}a_9\right)\;,\displaybreak[3]\nonumber\\
Z_{\chi_2}&=1+\varepsilon\left(a_{24}-a_1-a_7-a_{10}+2a_0-\left(\frac{\chi_1}{\chi_2}a_8+\frac{\chi_4}{\chi_2}a_9\right)\right)\;,\displaybreak[3]\nonumber\\
Z_{\chi_3}&=1+\varepsilon\left(a_{25}-a_1-a_7-a_{10}+2a_0-\left(\frac{\chi_1}{\chi_3}a_8+\frac{\chi_4}{\chi_3}a_9\right)\right)\;,\displaybreak[3]\nonumber\\
Z_{\chi_4}&=1+\varepsilon\left(a_{26}-a_1-2a_{10}+2a_0-\frac{\chi_2+\chi_3}{\chi_4}a_8\right)\;,\displaybreak[3]\nonumber\\
Z_{\varrho_1}&=1+\varepsilon\left(a_{27}-a_1-2a_7+2a_0-\frac{\varrho_2+\varrho_3}{\varrho_1}a_9\right)\;,\displaybreak[3]\nonumber\\
Z_{\varrho_2}&=1+\varepsilon\left(a_{28}-a_1-a_7-a_{10}+2a_0-\left(\frac{\varrho_1}{\varrho_2}a_8+\frac{\varrho_4}{\varrho_2}a_9\right)\right)\;,\displaybreak[3]\nonumber\\
Z_{\varrho_3}&=1+\varepsilon\left(a_{29}-a_1-a_7-a_{10}+2a_0-\left(\frac{\varrho_1}{\varrho_3}a_8+\frac{\varrho_4}{\varrho_3}a_9\right)\right)\;,\displaybreak[3]\nonumber\\
Z_{\varrho_4}&=1+\varepsilon\left(a_{30}-a_1-2a_{10}+2a_0-\frac{\varrho_2+\varrho_3}{\varrho_4}a_8\right)\;,\displaybreak[3]\nonumber\\
Z_{\zeta_1}&=1+\varepsilon\left(a_{31}-4a_7+4a_0-\frac{\zeta_2+\zeta_3+\zeta_5+\zeta_9}{\zeta_1}a_9\right)\;,\displaybreak[3]\nonumber\\
Z_{\zeta_2}&=1+\varepsilon\left(a_{32}-3a_7-a_{10}+4a_0-\left(\frac{\zeta_1}{\zeta_2}a_8+\frac{\zeta_4+\zeta_6+\zeta_{10}}{\zeta_2}a_9\right)\right)\;,\displaybreak[3]\nonumber\\
Z_{\zeta_3}&=1+\varepsilon\left(a_{33}-3a_7-a_{10}+4a_0-\left(\frac{\zeta_1}{\zeta_3}a_8+\frac{\zeta_4+\zeta_7+\zeta_{11}}{\zeta_3}a_9\right)\right)\;,\displaybreak[3]\nonumber\\
Z_{\zeta_4}&=1+\varepsilon\left(a_{34}-2a_7-2a_{10}+4a_0-\left(\frac{\zeta_2+\zeta_3}{\zeta_4}a_8+\frac{\zeta_8+\zeta_{12}}{\zeta_4}a_9\right)\right)\;,\displaybreak[3]\nonumber\\
Z_{\zeta_5}&=1+\varepsilon\left(a_{35}-2a_7-2a_{10}+4a_0-\left(\frac{\zeta_1}{\zeta_5}a_8+\frac{\zeta_6+\zeta_7+\zeta_{13}}{\zeta_5}a_9\right)\right)\;,\displaybreak[3]\nonumber\\
Z_{\zeta_6}&=1+\varepsilon\left(a_{36}-2a_7-2a_{10}+4a_0-\left(\frac{\zeta_2+\zeta_5}{\zeta_6}a_8+\frac{\zeta_8+\zeta_{14}}{\zeta_6}a_9\right)\right)\;,\displaybreak[3]\nonumber\\
Z_{\zeta_7}&=1+\varepsilon\left(a_{37}-2a_7-2a_{10}+4a_0-\left(\frac{\zeta_3+\zeta_5}{\zeta_7}a_8+\frac{\zeta_8+\zeta_{15}}{\zeta_7}a_9\right)\right)\;,\displaybreak[3]\nonumber\\
Z_{\zeta_8}&=1+\varepsilon\left(a_{38}-a_7-3a_{10}+4a_0-\left(\frac{\zeta_4+\zeta_6+\zeta_7}{\zeta_8}a_8+\frac{\zeta_{16}}{\zeta_8}a_9\right)\right)\;,\displaybreak[3]\nonumber\\
Z_{\zeta_9}&=1+\varepsilon\left(a_{39}-3a_7-a_{10}+4a_0-\left(\frac{\zeta_1}{\zeta_9}a_8+\frac{\zeta_{10}+\zeta_{11}+\zeta_{13}}{\zeta_9}a_9\right)\right)\;,\displaybreak[3]\nonumber\\
Z_{\zeta_{10}}&=1+\varepsilon\left(a_{40}-2a_7-2a_{10}+4a_0-\left(\frac{\zeta_2+\zeta_9}{\zeta_{10}}a_8+\frac{\zeta_{12}+\zeta_{14}}{\zeta_{10}}a_9\right)\right)\;,\displaybreak[3]\nonumber\\
Z_{\zeta_{11}}&=1+\varepsilon\left(a_{41}-2a_7-2a_{10}+4a_0-\left(\frac{\zeta_3+\zeta_9}{\zeta_{11}}a_8+\frac{\zeta_{12}+\zeta_{15}}{\zeta_{11}}a_9\right)\right)\;,\displaybreak[3]\nonumber\\
Z_{\zeta_{12}}&=1+\varepsilon\left(a_{42}-a_7-3a_{10}+4a_0-\left(\frac{\zeta_{14}+\zeta_{10}+\zeta_{11}}{\zeta_{12}}a_8+\frac{\zeta_{16}}{\zeta_{12}}a_9\right)\right)\;,\displaybreak[3]\nonumber\\
Z_{\zeta_{13}}&=1+\varepsilon\left(a_{43}-2a_7-2a_{10}+4a_0-\left(\frac{\zeta_5+\zeta_9}{\zeta_{13}}a_8+\frac{\zeta_{14}+\zeta_{15}}{\zeta_{13}}a_9\right)\right)\;,\displaybreak[3]\nonumber\\
Z_{\zeta_{14}}&=1+\varepsilon\left(a_{44}-a_7-3a_{10}+4a_0-\left(\frac{\zeta_6+\zeta_{10}+\zeta_{13}}{\zeta_{14}}a_8+\frac{\zeta_{16}}{\zeta_{14}}a_9\right)\right)\;,\displaybreak[3]\nonumber\\
Z_{\zeta_{15}}&=1+\varepsilon\left(a_{45}-a_7-3a_{10}+4a_0-\left(\frac{\zeta_7+\zeta_{11}+\zeta_{13}}{\zeta_{15}}a_8+\frac{\zeta_{16}}{\zeta_{15}}a_9\right)\right)\;,\displaybreak[3]\nonumber\\
Z_{\zeta_{16}}&=1+\varepsilon\left(a_{46}-4a_{10}+4a_0-\frac{\zeta_8+\zeta_{12}+\zeta_{14}+\zeta_{15}}{\zeta_{16}}a_8\right)\;,\displaybreak[3]\nonumber\\
Z_{\vartheta_1}&=1+\varepsilon\left(a_{47}-a_1-4a_7+5a_0-\frac{\vartheta_2+\vartheta_3+\vartheta_5+\vartheta_9}{\vartheta_1}a_9\right)\;,\displaybreak[3]\nonumber\\
Z_{\vartheta_2}&=1+\varepsilon\left(a_{48}-a_1-3a_7-a_{10}+5a_0-\left(\frac{\vartheta_1}{\vartheta_2}a_8+\frac{\vartheta_4+\vartheta_6+\vartheta_{10}}{\vartheta_2}a_9\right)\right)\;,\displaybreak[3]\nonumber\\
Z_{\vartheta_3}&=1+\varepsilon\left(a_{49}-a_1-3a_7-a_{10}+5a_0-\left(\frac{\vartheta_1}{\vartheta_3}a_8+\frac{\vartheta_4+\vartheta_7+\vartheta_{11}}{\vartheta_3}a_9\right)\right)\;,\displaybreak[3]\nonumber\\
Z_{\vartheta_4}&=1+\varepsilon\left(a_{50}-a_1-2a_7-2a_{10}+5a_0-\left(\frac{\vartheta_2+\vartheta_3}{\vartheta_4}a_8+\frac{\vartheta_8+\vartheta_{12}}{\vartheta_4}a_9\right)\right)\;,\displaybreak[3]\nonumber\\
Z_{\vartheta_5}&=1+\varepsilon\left(a_{51}-a_1-2a_7-2a_{10}+5a_0-\left(\frac{\vartheta_1}{\vartheta_5}a_8+\frac{\vartheta_6+\vartheta_7+\vartheta_{13}}{\vartheta_5}a_9\right)\right)\;,\displaybreak[3]\nonumber\\
Z_{\vartheta_6}&=1+\varepsilon\left(a_{52}-a_1-2a_7-2a_{10}+5a_0-\left(\frac{\vartheta_2+\vartheta_5}{\vartheta_6}a_8+\frac{\vartheta_8+\vartheta_{14}}{\vartheta_6}a_9\right)\right)\;,\displaybreak[3]\nonumber\\
Z_{\vartheta_7}&=1+\varepsilon\left(a_{53}-a_1-2a_7-2a_{10}+5a_0-\left(\frac{\vartheta_3+\vartheta_5}{\vartheta_7}a_8+\frac{\vartheta_8+\vartheta_{15}}{\vartheta_7}a_9\right)\right)\;,\displaybreak[3]\nonumber\\
Z_{\vartheta_8}&=1+\varepsilon\left(a_{54}-a_1-a_7-3a_{10}+5a_0-\left(\frac{\vartheta_4+\vartheta_6+\vartheta_7}{\vartheta_8}a_8+\frac{\vartheta_{16}}{\vartheta_8}a_9\right)\right)\;,\displaybreak[3]\nonumber\\
Z_{\vartheta_9}&=1+\varepsilon\left(a_{55}-a_1-3a_7-a_{10}+5a_0-\left(\frac{\vartheta_1}{\vartheta_9}a_8+\frac{\vartheta_{10}+\vartheta_{11}+\vartheta_{13}}{\vartheta_9}a_9\right)\right)\;,\displaybreak[3]\nonumber\\
Z_{\vartheta_{10}}&=1+\varepsilon\left(a_{56}-a_1-2a_7-2a_{10}+5a_0-\left(\frac{\vartheta_2+\vartheta_9}{\vartheta_{10}}a_8+\frac{\vartheta_{12}+\vartheta_{14}}{\vartheta_{10}}a_9\right)\right)\;,\displaybreak[3]\nonumber\\
Z_{\vartheta_{11}}&=1+\varepsilon\left(a_{57}-a_1-2a_7-2a_{10}+5a_0-\left(\frac{\vartheta_3+\vartheta_9}{\zeta_{11}}a_8+\frac{\vartheta_{12}+\vartheta_{15}}{\vartheta_{11}}a_9\right)\right)\;,\displaybreak[3]\nonumber\\
Z_{\vartheta_{12}}&=1+\varepsilon\left(a_{58}-a_1-a_7-3a_{10}+5a_0-\left(\frac{\vartheta_{14}+\vartheta_{10}+\vartheta_{11}}{\vartheta_{12}}a_8+\frac{\vartheta_{16}}{\vartheta_{12}}a_9\right)\right)\;,\displaybreak[3]\nonumber\\
Z_{\vartheta_{13}}&=1+\varepsilon\left(a_{59}-a_1-2a_7-2a_{10}+5a_0-\left(\frac{\vartheta_5+\vartheta_9}{\vartheta_{13}}a_8+\frac{\vartheta_{14}+\vartheta_{15}}{\vartheta_{13}}a_9\right)\right)\;,\displaybreak[3]\nonumber\\
Z_{\vartheta_{14}}&=1+\varepsilon\left(a_{60}-a_1-a_7-3a_{10}+5a_0-\left(\frac{\vartheta_6+\vartheta_{10}+\vartheta_{13}}{\vartheta_{14}}a_8+\frac{\vartheta_{16}}{\vartheta_{14}}a_9\right)\right)\;,\displaybreak[3]\nonumber\\
Z_{\vartheta_{15}}&=1+\varepsilon\left(a_{61}-a_1-a_7-3a_{10}+5a_0-\left(\frac{\vartheta_7+\vartheta_{11}+\vartheta_{13}}{\vartheta_{15}}a_8+\frac{\vartheta_{16}}{\vartheta_{15}}a_9\right)\right)\;,\displaybreak[3]\nonumber\\
Z_{\vartheta_{16}}&=1+\varepsilon\left(a_{62}-a_1-4a_{10}+5a_0-\frac{\vartheta_8+\vartheta_{12}+\vartheta_{14}+\vartheta_{15}}{\vartheta_{16}}a_8\right)\;.
\label{ren7}
\end{align}

\section{Alternative renormalization of the parameters}\label{Pr}

It was presented in Section \ref{STABILITY} the renormalization of the coefficients related to the mass parameters, vertices and vacuum terms. An alternative, but equivalent, way to present the renormalization of the dimensionless coefficients (Appendix \ref{App2}) can be performed by using the matricial renormalization. This happens due the fact that the mixing between the quantum sources induces, in a natural way, a mixing between their respective parameters. Thus, we can simply write
\begin{eqnarray}
\begin{pmatrix}
\alpha_{01}\\
\alpha_{02}\\
\alpha_{03}\\
\alpha_{04}\\
\alpha_{05}\\
\end{pmatrix}=\mathcal{Z}_{\alpha}\begin{pmatrix}
\alpha_1\\
\alpha_2\\
\alpha_3\\
\alpha_4\\
\alpha_5\\
\end{pmatrix}\;.
\label{renm1xx}
\end{eqnarray}
It is found that
\begin{equation}
\mathcal{Z}_\alpha =\mathbb{1}+\varepsilon\begin{pmatrix}
a_{11}-2a_7+a_0& -a_9&-a_9&0\\
-a_8&a_{12}-a_7-a_{10}+a_0&0&-a_9\\
-a_8&0&a_{13}-a_7-a_{10}+a_0&-a_9\\
0&-a_8&-a_8&a_{14}-2a_{10}+a_0
\end{pmatrix}\;.
\end{equation}
And it is a straightforward exercise to generalize the method to the other classes of parameters.

\section*{Acknowledgements}

We acknowledge A.~A.~Tomaz for the help with the propagators computation. The Conselho Nacional de Desenvolvimento Cient\'{i}fico e Tecnol\'{o}gico\footnote{RFS is a PQ-2 level researcher under the program \emph{Produtividade em Pesquisa}, 308845/2012-9.} (CNPq-Brazil), The Coordena\c c\~ao de Aperfei\c coamento de Pessoal de N\'ivel Superior (CAPES) and the Pr\'o-Reitoria de Pesquisa, P\'os-Gradua\c c\~ao e Inova\c c\~ao (PROPPI-UFF) are acknowledge for financial support.

\end{document}